\documentclass{article}
\usepackage[dvips,dvipdfm]{graphicx}
\usepackage{longtable}

\begin{document}
\setlength{\baselineskip}{24pt}

%\begin{frontmatter}
%\title{Study of 23 day periodicity of Blazar Mkn501 in 1997}
%\author[AIST]{S.Osone}
%\address[AIST]{Advance Industrial Science and Technology, Tsukuba central 2, Tsukuba, 305-8568, Japan, osone@icrr.u-tokyo.ac.jp}

{\bf Study of 23 day periodicity of Blazar Mkn501 in 1997}

S.Osone

Advance Industrial Science and Technology, Higashi Central 5, Tsukuba, 305-8565, Japan, osone@icrr.u-tokyo.ac.jp

\section{Abstract}
We confirm a 23 day periodicity during a large flare in 1997 for X-ray data of X-ray satellite {\it RXTE} all sky monitor (ASM), two TeV gamma ray data from Utah Seven Telescope and HEGRA, with a Fourier analysis.
We found the three results to be the same with a newly estimated error. We confirm the presence of a frequency dependent power (1/f noise) in a frequency-power diagram. Further, we calculated a chance probability of the occurrence of the 23 day periodicity by considering the 1/f noise and obtained a chance probability 4.88 $\times10^{-3}$ for the HEGRA data:this is more significant than the previous result by an order. We also obtained an indentical periodicity with another kind of timing analysis--epoch folding method for the ASM data and HEGRA data. We strongly suggest an existence of the periodicity.
We divided the HEGRA data into two data sets, analyzed them with a Fourier method, and found an unstableness of the periodicity with a 3.4 sigma significance.
We also analyzed an energy spectra of the X-ray data of a {\it RXTE} proportional counter array and we found that a combination of three physical parameters-- a magnetic field, a Lorentz factor, and a beaming factor--is related to the periodicity.

%\begin{keyword}
%PACS:95.85.Nv;95.85.Pw;98.54.Cm
%\end{keyword}

%\end{frontmatter}

\section{Introduction}
Gamma ray emission from active galactic nuclei (AGN) has been measured by detectors in an orbit and by detectors on the ground.
The EGRET~\cite{Thompson} detector on the gamma ray satellite {\it CGRO} was sensitive to GeV gamma rays and detected 90 Blazars (e.g.,~\cite{Hartman}).
 Air Cherenkov detectors on the ground are sensitive to TeV energies.
They detected the ten Blazars--Mkn 421 (z = 0.030), Mkn 501 (z = 0.034), PKS 2155-304 (z = 0.116), 1ES1959+650 (z = 0.048), 1ES 2344+514 (z = 0.044), H1426+428(z = 0.129), PKS2005-489 (z = 0.071), H2356-309 (z = 0.165), 1ES1218+304 (z = 0.182) and 1ES1101-232 (z = 0.186).
The  energy spectra of Blazars has two components.
One component extends from the radio to the X-ray band, while the other is in the gamma ray range.
Low energy photons are interpreted to be a result of synchrotron emission by accelerated high energy electrons, and high energy photons are interpreted to be resulted of inverse Compton scattering of the synchrotron emission by high energy electrons. 
There are two types of Blazars, the BL Lac type and the QSO type.
The BL Lac is of two types: high frequency BL Lac (HBL) and low frequency BL Lac (LBL), where the names correspond to the peak frequency of the synchrotron emission. The peak frequency of the synchrotron emission and that of the inverse Compton emission occur in the radio band  and  Xray band for LBL and in the X-ray band and TeV band for HBL.

The TeV gamma ray flux of HBLs Mkn421 and Mkn501 are usually lower than that of the Crab nebula.
In 1997, the flux of Mkn501 increased up to 10 Crab over 3 months. Two Cherenkov detectors--Utah Seven Telescope (Utah TA) and HEGRA--and X-ray satellite {\it RXTE} all sky monitor (ASM) observed Mkn501 simultaneously during this flare.
Hayashida et al.~\cite{UtahTA} studied the Utah TA data during this flare with a Fourier analysis and suggested two periodicities--13 day and 23 day.
Kranich~\cite{Dron} studied both the HEGRA and ASM data during this flare with a Fourier analysis and obtained a 22.5 day periodicity with a chance probability of 0.028 for the HEGRA data and a 22.5 day periodicity with a chance probability of 0.047 for the ASM data.
These results weakly suggest the 23 day periodicity.
There is some problem in these results.
The  frequency-power diagram shows frequency dependent power (1/f noise).
This 1/f noise in the frequency-power diagram is well known in AGN and a Blackhole candidate binary in X-ray band. The origin of the noise is unknown.
Hayashida et al.~\cite{UtahTA} do not consider an effect of the 1/f noise on both a significance and an error of the periodicity.
Kranich~\cite{Dron} does not consider an effect of the 1/f noise on an error of the periodicity, but he considers an effect of the 1/f noise on a significance of the periodicity.
However, he uses an unreliable model of the 1/f noise for deducing a chance probability. He fitted a raw power spectra to obtain a model of the 1/f noise without discussing model reliability. With Poisson statistics, the power in a raw power spectra has a 100\% error because it follows a $\chi^2$ distribution of 2 degrees of freedom. In the case of Poisson statistics and the 1/f noise, this error is less than 100\%:however, power still has a large error.

In this study, we study these three data sets with a Fourier analysis and also use another kind of a timing analysis in order to increase the reliability of the periodicity.
We binned a raw power spectra in order to obtain a power spectra that is statistically reasonable, obtain the best model of the 1/f noise, and  estimate a chance probability by considering the reliability of the 1/f noise model. We obtain a lower chance probability than Kranich~\cite{Dron} by an order.
We obtain an error of the periodicity by considering an effect of the 1/f noise and found that the three results are same with a newly estimated error.
We study the stability of the periodicity and analyze  an energy spectra of the X-ray satellite {\it RXTE} proportional counter array (PCA) during this flare , in order to set a limit on an origin of the periodicity.

\section{data}
\subsection{timing analysis}
We obtained the HEGRA data~\cite{Dron} in 1997 as a form of (a MJD, flux, an error of flux). There are four kinds of data--CT1 (no moon), CT1(moon), CT2, and CTsys corresponding to different data acquisition conditions or detectors.
We use summed data for a timing analysis and show a lightcurve of the HEGRA data in MJD 50545-50661 in figure 1.
We obtained the Utah TA data~\cite{UtahTA} in 1997 as the form of (a MJD, flux, an error of flux) for a timing analysis and show a lightcurve of the Utah TA data in figure 2.
We obtained the ASM data of a 90 s dwell in the form of (a MJD, a rate, an error of a rate) for a timing analysis. The error of a rate for the ASM data includes a systematic error of 3\% derived using a lightcurve of Crab.
We show a lightcurve of the ASM data from 1996 to 2000 in figure 3. 
We use a span MJD 50545-50661 for both the HEGRA and ASM data, which is the same as in Kranich~\cite{Dron}. Further we use a span MJD 50520-50665 for the Utah TA data.
We show a lightcurve of the ASM data in MJD 50545-50661 in figure 4. 
We also use a span MJD 50300-50900 for the ASM data because the ASM data is plentiful and we require more than 10 cycles for increasing a reliability of a periodicity.

\subsection{spectral analysis}
We obtain the PCA raw data--standard 2 data that is suitable for an energy spectral analysis--and use a span MJD 50300-50900.
We use only the data of the top layer in the PCA, which has a low background. 
Using a standard tool FTOOLS 4.2, we selected the data with normal conditions--an elevation greater than 10 degrees, neglecting time of the SAA passage and time less than 30 minuites of the SAA passage, offset $\le 0.02$, and electron0 $\le 0.1$. We use a background model, L7 model, which is generally used for a faint source.
 We obtain one energy spectra for each continuous observation for 10 ksec.
The total number of energy spectra is 56.
We calibrate a systematic error of the PCA data with an energy spectra of Crab.
We fitted the energy spectra of Crab with an absorbed power law of the form dN/dE = $e^{-N_{\rm H}\sigma(E)} E^{-\alpha}$ and obtain a statistical disagreement $\chi^2$/d.o.f$ = $26.35. Here, $\sigma(E)$ is a crosssection of a photoabsorption, and $\alpha$ is a photon index.
We use a column density of neutral hydrogen $N_{\rm H}=32.70\times10^{20}$ cm$^{-2}$, which is given by EINLINE.
When we add a 1\% systematic error to the energy spectra of Crab and fitted this with an abosorbed power law, we obtain a statistical agreement $\chi^2$/d.o.f = 1.25.
The energy spectra of Crab should be fitted with an absorbed power law. We need 1 \% systematic error in order to fit the energy spectra of Crab with an absorbed power law. We found that there is 1 \% systematic error in PCA data itself.
Therefore, we add 1\% systematic error to an energy spectra of Mkn501.
We fitted an energy spectra of Mkn501 with an absorbed power law of a form dN/dE$=e^{-N_{\rm H}\sigma(E)} E^{-\alpha}$. We use $N_{\rm H}=$ 2.08 $\times10^{20}$ cm$^{-2}$, which is given by EINLINE.
We obtain a data set in the form of (a MJD, a photon index, an error of a photon index).

\section{timing analysis}
\subsection{periodicity}
When we use a normal Fourier analysis for an unevenly spaced data set, leakage of the power to neighbouring frequencies is a problem. A window function is normally applied to each data point.
We instead use a Fourier analysis in which a weighted power is applied~\cite{Scargle}~\cite{Numerical}.
The formula is as follows.

\begin{equation}
P(w) = \frac{1}{2\sigma^2}\Bigl[ \frac{ [\sum_j (h_j - \bar{h}){\rm cos}w(t_j - \tau)]^2 }{ \sum_j {\rm cos}^2w(t_j - \tau ) } +\frac{ [\sum_j ( h_j - \bar{h}){\rm sin}w(t_j - \tau)]^2}{ \sum_j {\rm sin}^2w(t_j - \tau)}\Bigl] 
\end{equation}

\begin{equation}
{\rm tan}(2w\tau) =\frac{ \sum_j {\rm sin}2wt_j }{ \sum_j {\rm cos}2wt_j }
\end{equation}

Here, $\bar{h}$ is an average of a count rate, $\sigma$ is variance of data, ($t_i, h_i$) are the $i$th observation time and rate respectively, and $\tau$ is an input that does not vary with a time offset.
This formula is the same with a least square analysis~\cite{Lomb}.
Kranich~\cite{Dron} used the same method.
We calculate a power spectra with 100$\times N/2$ frequencies in order to compensate a gap in a frequency and increase the reliability of a power spectra because a power has an error of about 100\% as  previously mentioned in the introduction.
$N$ is the number of data and $N/2$ is the number of independent frequencies.
We show the power spectra for three data sets in figure 5.
We obtain a maximum power as a periodicity.
We obtain a 22.5 day periodicity (5$\times10^{-7}$Hz) for the HEGRA and ASM data. We obtain a 23.6 day periodicity as the second largest peak for the Utah TA data. We obtain a peak around second harmonics (1$\times10^{-6}$ Hz) in power spectra of Utah TA and ASM data. We do not obtain a peak around secound harmonics in power spectra of HEGRA data. 
We also obtain a 23.6 day periodicity for the ASM data in MJD 50300-50900.
We discuss an error of these periodicities at a later stage.

We also attempt to another timing analysis--epoch folding method--in order to increase the reliability of the periodicity for the HEGRA and ASM data. 
We make a folded lightcurve with a period $P$ from 1 day to 116 days in steps of 0.5 day, and calculate $\chi^2(P)$=$\Sigma (h_i - h_0)^2/\sigma_i$ and $\chi^2(P)$/d.o.f. for each folded lightcurve. Here, $h_i$ is a rate of the ith data points, $h_0$ is an average count rate, and $\sigma_i$ is an error of the ith data points in a folded lightcurve.
We obtain a maximum peak in $\chi^2(P)$/d.o.f vs a period $P$ as a periodicity.
The lightcurve is composed of  multiple frequencies.
The epoch folding analysis is sensitive to the sum of multiple frequencies although a Fourier analysis is sensitive to a monochromatic frequency.
We show $\chi^2(P)$/d.o.f diagrams in figure 6.
A 22.5 day periodicity is present in the ASM data.
A 22.5 day periodicity is present as the second largest peak in the HEGRA data.
We found an identical periodicity with an epoch folding method.
We show the phase diagrams for a 22.5 day periodicity of the HEGRA and ASM data in figure 7. We confirm that the phase diagrams are broad.

A power spectra has noise equal to 1/f.
When only Poission statistics exist, we can estimate a chance probability of a detected periodicity with exp($-z$). Here, $z$ is a power. 
When there are Poisson statistics and 1/f noise, the chance probability does not follow  $\exp(-z)$. 
We have to calculate a chance probability by making simulated data sets that show the 1/f noise in a power spectra.
First, we must obtain a model of the 1/f noise for each data set.
We calculate a power spectra with N/2 frequencies for a fourier analysis in order to obtain independent data points.
We make a binned power spectra in order to obtain power spectra that is statistically reasonable as shown in figure 8. We confirm the 1/f noise for three data sets.
We fit these power spectra with a model of a form, power $P(f)=1+\alpha \times f^{-\beta}$. 
Here, $f$ is a frequency and $\alpha, \beta$ are constants.
$P(f)=$1 indicates Poisson statistics. We note an area at $10^{-5}$ Hz and $10^{-7}$ Hz as A.
We obtain (A,$\beta$) instead of ($\alpha, \beta$) because $\alpha$ and $\beta$ are strongly coupled and have a large error.
We show a fitted power spectra with a model in figure 8 and fitted parameters in table 1.
We remove the data point at 5$\times10^{-7}$ Hz as a periodicity when we fit the HEGRA data although this data point is included in figure 8.
Second, we create 1000 simulated data sets in the form of (a MJD, a rate) by taking an inverse Fourier transform of this model ($f, P(f)$)~\cite{Takeshima}. 
We input two parameters (mod, $\beta$') to generate the simulated data. Here, a mod is a ratio parameter between the Poisson statistics and 1/f noise.
We use an actual time history for the simulation data.
We analyze the 1000 simulation data sets with a Fourier method, take an average of the 1000 powers in each frequency, and make one binned power spectra. We then fit this power spectra with a model of a form, power $P(f)=1+\alpha \times f^{-\beta}$, and obtain ($A',\beta$''). $\beta$' and $\beta$'' are not always the same.
We consider two extreme conditions (A$+\Delta$A, $\beta-\Delta \beta$)(case 1) and (A$+\Delta$A, $\beta+\Delta \beta$)(case 2), using which we obtain a low significance for the periodicity. Here, $\Delta$ is a 1 sigma statistical error.
We tune (mod, $\beta$') so that ($A',\beta$'') matches (A$+\Delta$A, $\beta-\Delta \beta$). We show simulated binned power spectra with adjusted parameters in figure 9 for the HEGRA data, figure 10 for the Utah TA data, and figure 11 for the ASM data.
Third, we create 10000 or 1000 simulation data sets with adjusted parameter values (mod, $\beta$'), analyze simulated data with a Fourier method, and search for a maximum power $P_{max}$ in each simulated power spectra. 
We show the distribution of maximum power $P_{max}$ in figure 9 for the HEGRA data, figure 10 for the Utah TA data, and figure 11 for the ASM data.
We calculate chance probability $P_{ch}=N(P_{max} \ge P_{obs})/N_0$.
Here, $P_{obs}$ is the power for the observed periodicity, and $N_0$ is the number of simulation data. 
We show $N(P_{max} \ge P_{obs})$, $N_0$ in table 1.
We perform the same analysis for another condition (A$+\Delta$A, $\beta+\Delta \beta$). 
We obtain lower chance probability for the two conditions as a chance probability of the periodicity. We obtain a chance probability of 4.88$\times10^{-3}$ for the HEGRA data, 0.981 for the Utah TA, and 0.200 for the ASM data, as shown in table 1.
The 22.5 day periodicity seems to be significant in the observed binned power spectra of HEGRA data as figure 8.
However, two fitting parameters $A, \beta$ of the 1/f noise have large errors as shown in table 1.
When we consider an extreme condition of case 1, 2, the simulated binned power spectra in these condition have large power as shown in figure 9.
Therefore, we obtain low chnace probability 4.88$\times10^{-3}$ for HEGRA data. For the epoch folding method, we do not obtain a significance of the periodicity because of complex treatment of huge data sets.

We estimate an error of the periodicity for a Fourier analysis. The chance probability does not follow exp($-z$) because there is an 1/f noise. Here, $z$ is a power. Therefore, we deduce an error of the periodicity with the simulation data.
We make the simulation data of Gaussain of (R, $\sigma$) in a form of (a MJD, a rate). Here, R is an observed count rate, and $\sigma$ is a variance of data.
The error of a count rate is only a Poisson fluctuation.
In order to include a fluctuation of the 1/f noise, we use a variance of the data.
We use an actual time history for the simulation data.
We create 1000 simulation data sets, make a Fourier analysis, search for a maximum peak as a periodicity in each power spectra and obtain 1000 periodicities.
We show the distributions of a periodicity for three data sets in figure 12.
We fitted each distribution of periodicities with a Gaussian and obtain a 1 sigma statistical error of the periodicity.
We found that the three periodicities are the same with a 1.3 sigma significance and found that a width of the periodicity is as narrow as $\Delta P/P \sim 0.01$.
In the case of the epoch folding method, we do not obtain an error of the periodicity because of complex treatment of huge data sets.

\subsection{stability of periodicity}
We study stability of the periodicity in order to set a limit on the origin of the periodicity. We divided statistically good data--HEGRA data--into two and analyzed these data with a Fourier method, as shown in table 2.
We estimated an error of a periodicity for a Fourier analysis, as described in the previous section:we show this in figure 13.
We found that the periodicity is unstable with a 3.4 sigma significance.

%The periodicity for epoch folding method is not same with that of a fourier method for two epochs.
%We consider this difference is attributed to a character of a timing analysis, fourier method is sensitive to a monochromatic frequency, epoch folding method to sum of multiple frequencies.
%This result means that both monochromatic and sum of multiple component change during two epochs.

\section{Change of energy spectra}
We create a phase diagram of a photon index for the 23.6 day periodicity, that is, a periodicity for the ASM data in MJD 50300-50900, as shown in figure 14.
A clear relation between a photon index and a phase can be observed.
X-rays have been considered as synchrotron emission by a synchrotron self Compton model.
There is an energy spectral change by a synchrotron cooling:however, the time scale is about $10^3$ s~\cite{Takahashi}.
Therefore, the observed relation is related to another physics.
A photon index $\alpha$ changes from 1.8 to 2.4. The index in an energy spectra $\nu F_{\nu}$(erg s$^{-1}$) vs. $\nu$ (Hz) is $-\alpha+2$. There is an index change from $-0.4$ to 0.2. 
We obtain a photon index 2.345$\pm$0.015 at a phase of 0.0 and found a negative index with (2.345-2.000)/0.015=23 sigma significance. We also obtain a photon index 1.874$\pm$0.0084 at a phase of 0.8 and found a positive index with (2.000-1.874)/0.0084 = 15 sigma significance. 
The PCA is sensitive to energy from 3 keV to 20 keV. A negative index means that a peak energy of the synchrotron emission is below 3 keV, and the positive index means that a peak energy of the synchrotron emission is above 20 keV. Therefore, a change in the index from a negative value to a positive value means that the peak energy of the synchrotron emission moves from a low energy to a high energy.
The peak energy of the synchrotron emission is written as $E_p=10^6 \gamma^2 B \delta /(1+z)$.
Here, $\gamma$ is a Lorentz factor of accelerated electrons, $B$ is a magnetic field, $\delta$ is a beaming factor of jet, and $z$ is a redshift.
We found that a combination of these parameters is related to the periodicity and changes during this flare by $\Delta(\gamma^2 B \delta)/(\gamma^2 B \delta) \ge 6$.

\section{Conclusion}
We analyze three data sets--HEGRA, Utah TA, and ASM--in 1997 for Mkn501 with a Fourier method and confirm a 23 day periodicity for these three data sets.
We found that the three results are the same with a newly estimated error.
We confirm a 1/f noise in a frequency-power diagram and obtain a chance probability of 4.88$\times10^{-3}$ for a periodicity in the HEGRA data by considering a 1/f noise: this is more significant than the previous result by an order.
We obtain the 23 day periodicity with more than 10 cycles for the ASM data.
We also obtain the same periodicity with an epoch folding method for the HEGRA and ASM data.
Therefore, we strongly suggest the existence of the 23 day periodicity.
We confirm that the phase diagrams of the HEGRA and ASM data are broad.
We found that a width of the periodicity is as narrow as $\Delta P/P \sim 0.01$.
We divided the HEGRA data into two, analyzed these data with a Fourier method, and found that a periodicity is unstable with a 3.4 sigma significance.
We analyzed an X-ray energy spectra of Mkn501 during this flare and found that a combination of three physical parameters--a magnetic field, a Lorentz factor, and a beaming factor--is related to the periodicity and is changed during this flare by $\Delta(\gamma^2 B \delta)/(\gamma^2 B \delta) \ge 6$.

We thank both Dr.Mitsuda, Dr.Dotani and Dr. Teshima for an useful advice and discussion across the complete study and analysis.
We thank Dr.Kranich for giving us the HEGRA data set.
We thank Dr.Yamaoka for the RXTE PCA energy spectra.

\begin{figure}[p]
\begin{center}
\includegraphics[width=12cm, height=6cm]{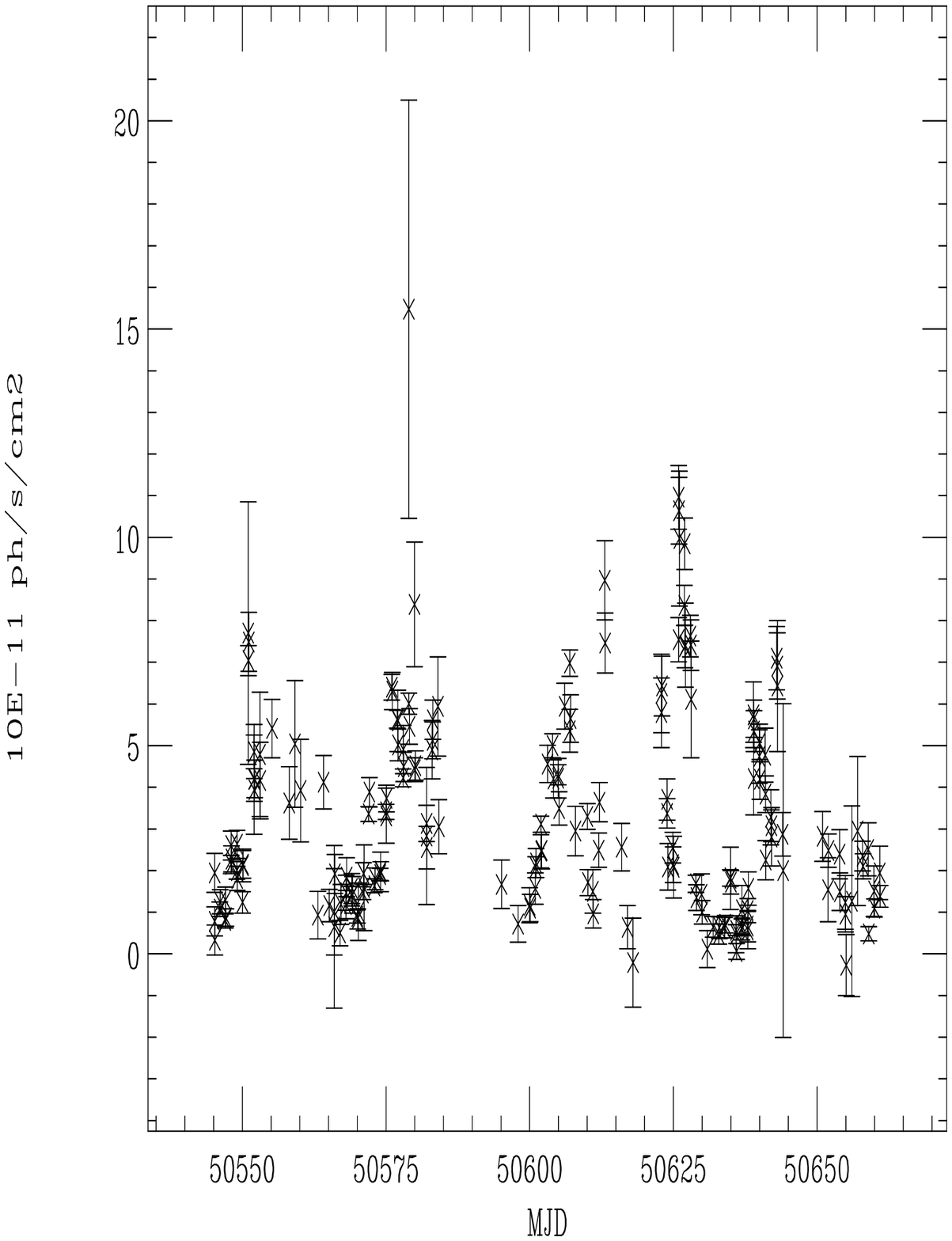}
\end{center}
\caption{A lightcurve of HEGRA data in MJD 50545-50661} 
\end{figure}

\begin{figure}[p]
\begin{center}
\includegraphics[width=12cm, height=6cm]{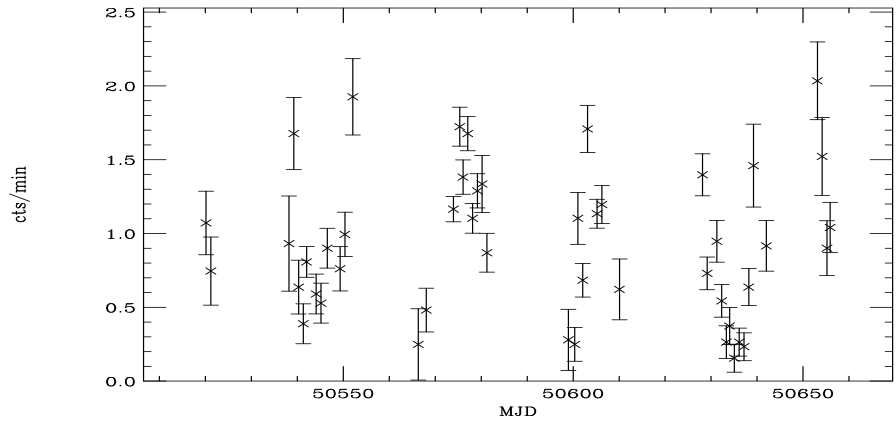}
\caption{A lightcurve of Utah TA in MJD 50520-50665}
\end{center}
\end{figure}

\begin{figure}[p]
\begin{center}
\includegraphics[width=12cm, height=6cm]{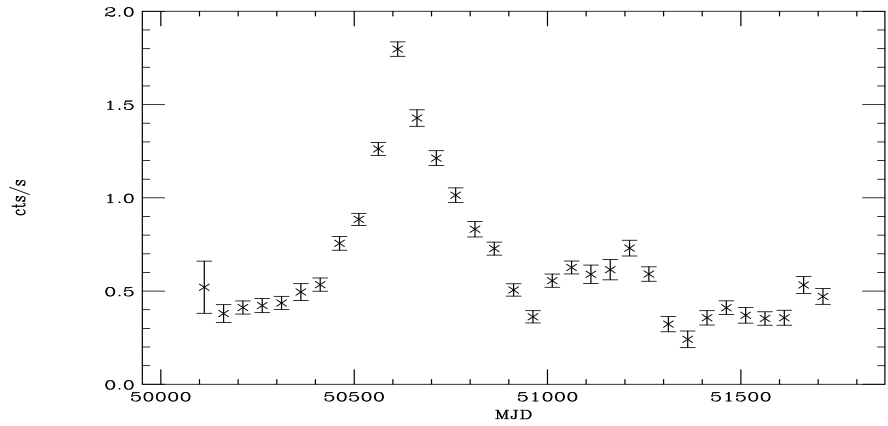}
\caption{A lightcurve of ASM data from 1996 to 2000. Data points are binned to a 50 day for clearness.}
\end{center}
\end{figure} 

\begin{figure}[p]
\begin{center}
\includegraphics[width=12cm, height=6cm]{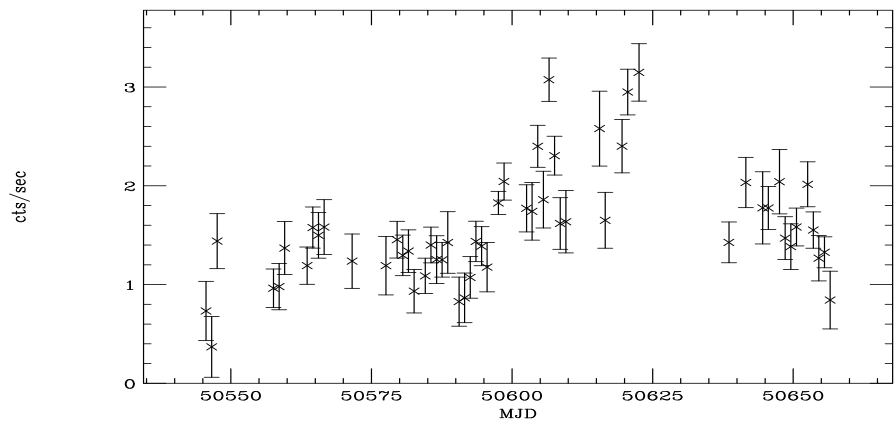}
\caption{A lightcurve of ASM data in MJD 50545-50661. Data points are binned to a 1 day for clearness and the binned count rate which includes data points above 20 are shown.}
\end{center}
\end{figure}

\begin{figure}[p]
 \begin{center}
\includegraphics[width=6cm, height=6cm]{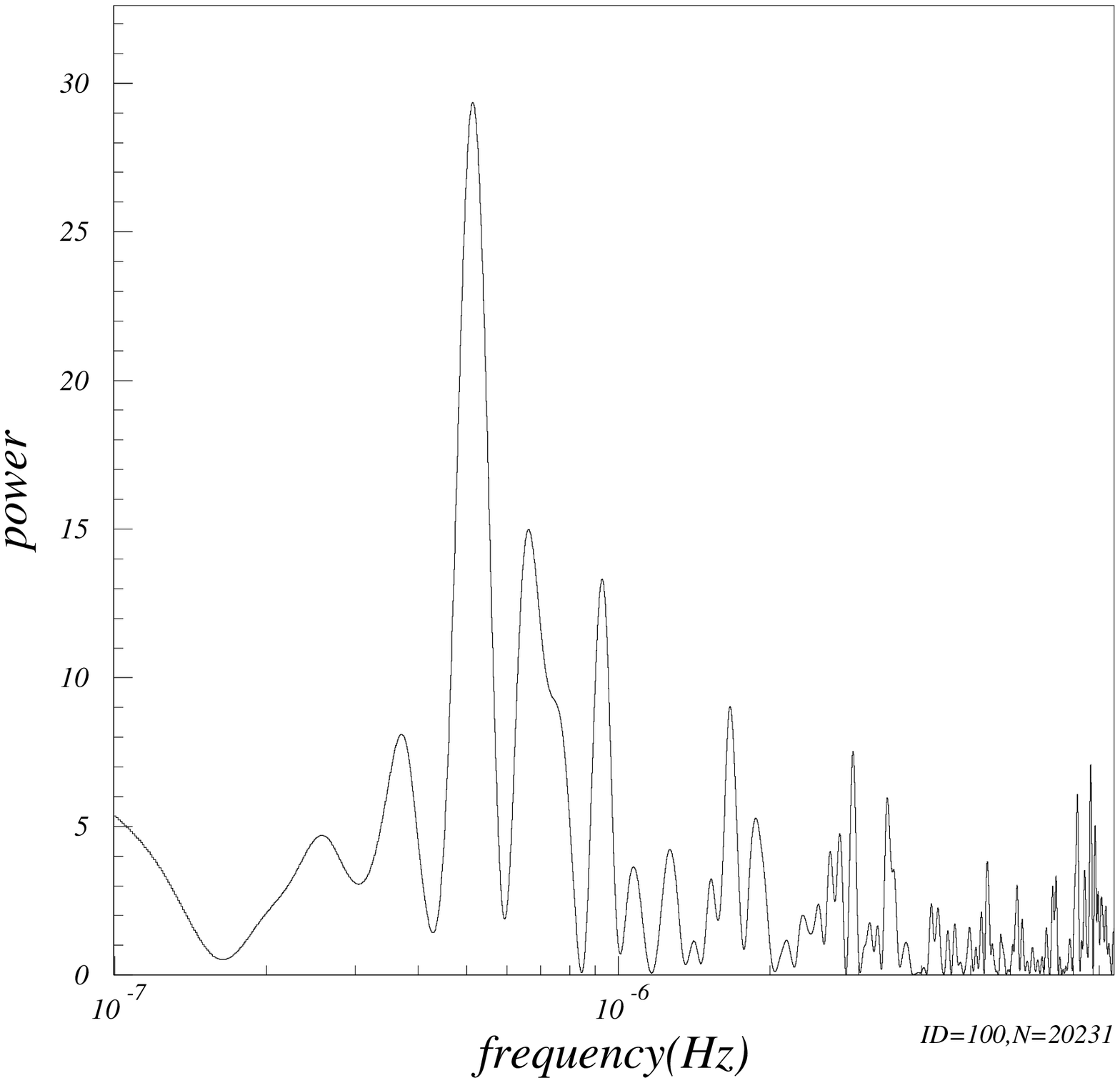}
\includegraphics[width=6cm, height=6cm]{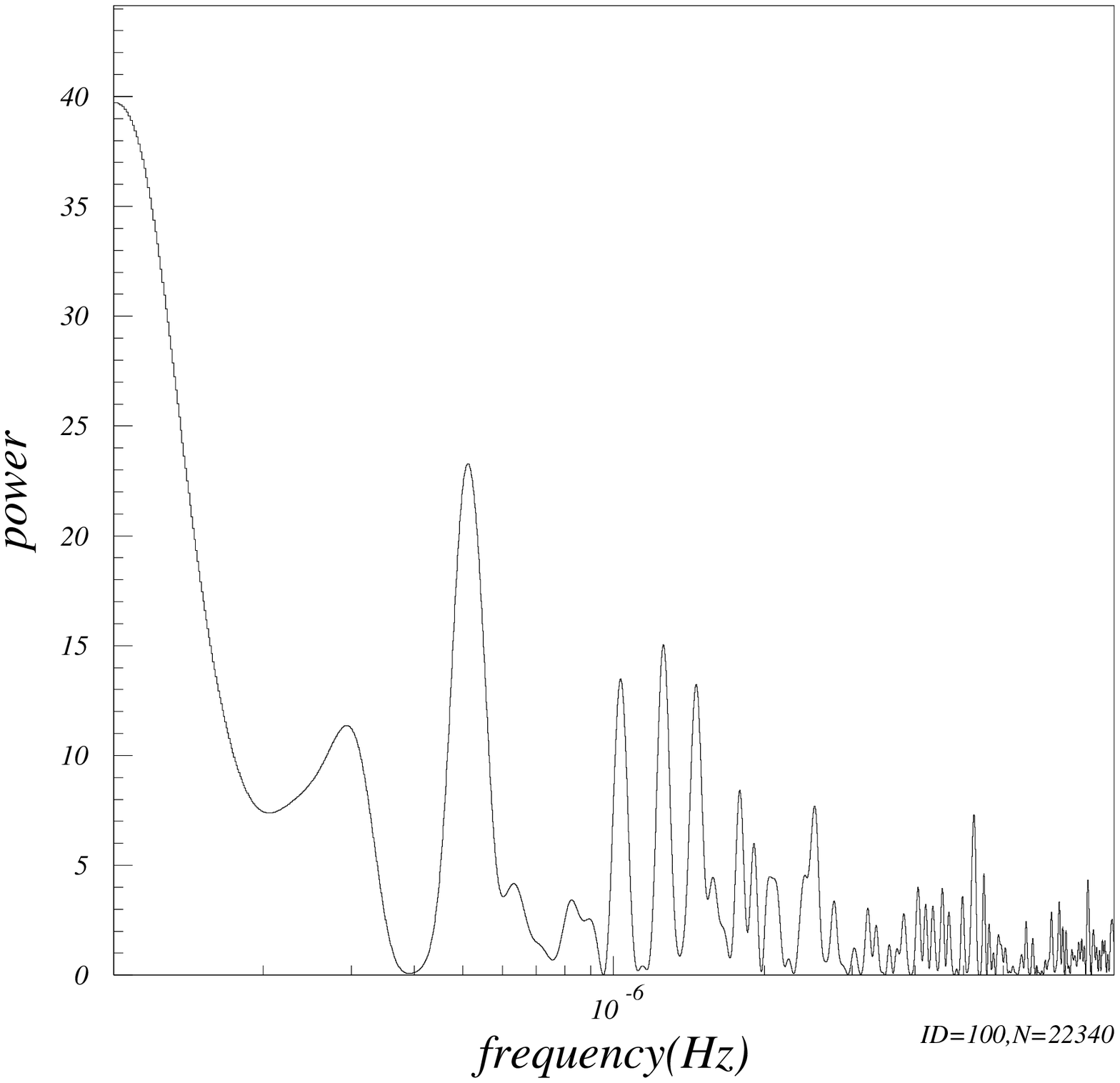}
\includegraphics[width=6cm, height=6cm]{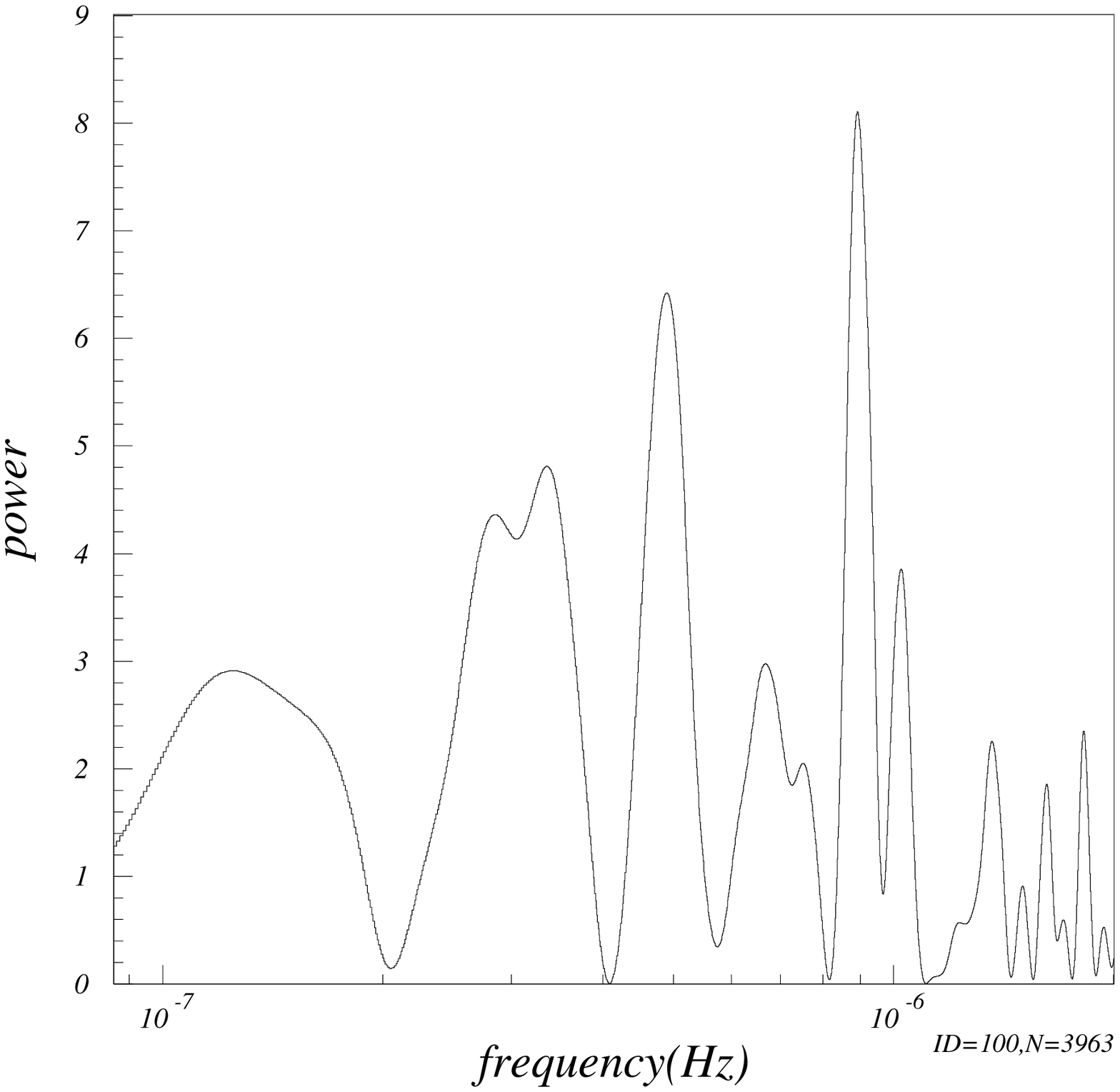}
 \end{center}
\caption{The power spectra with a fourier analysis for HEGRA data(top left), ASM data(top right) and Utah TA data(bottom)}
\end{figure}

\begin{figure}[p]
 \begin{center}
\includegraphics[width=6cm, height=6cm]{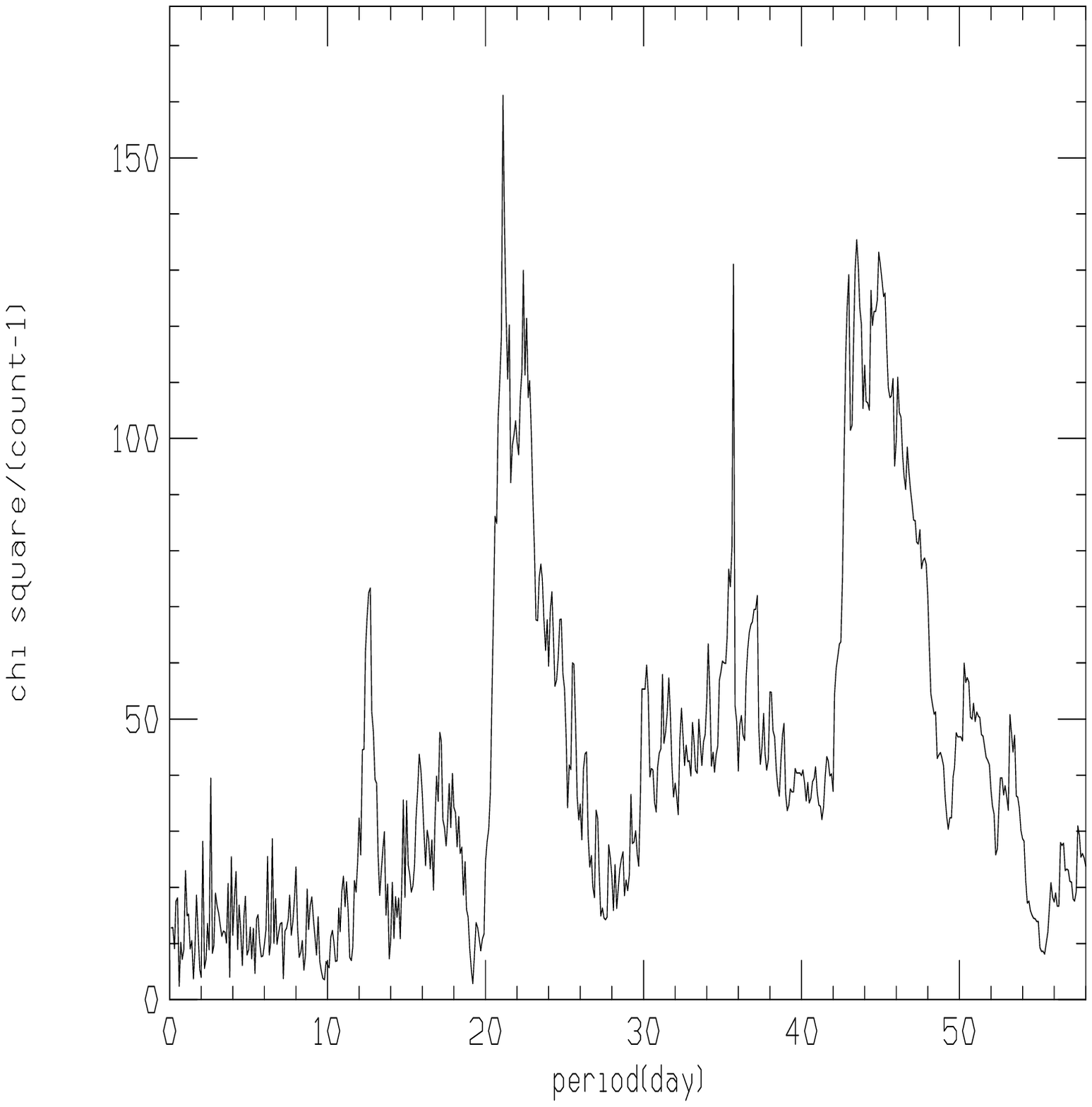}
\includegraphics[width=6cm, height=6cm]{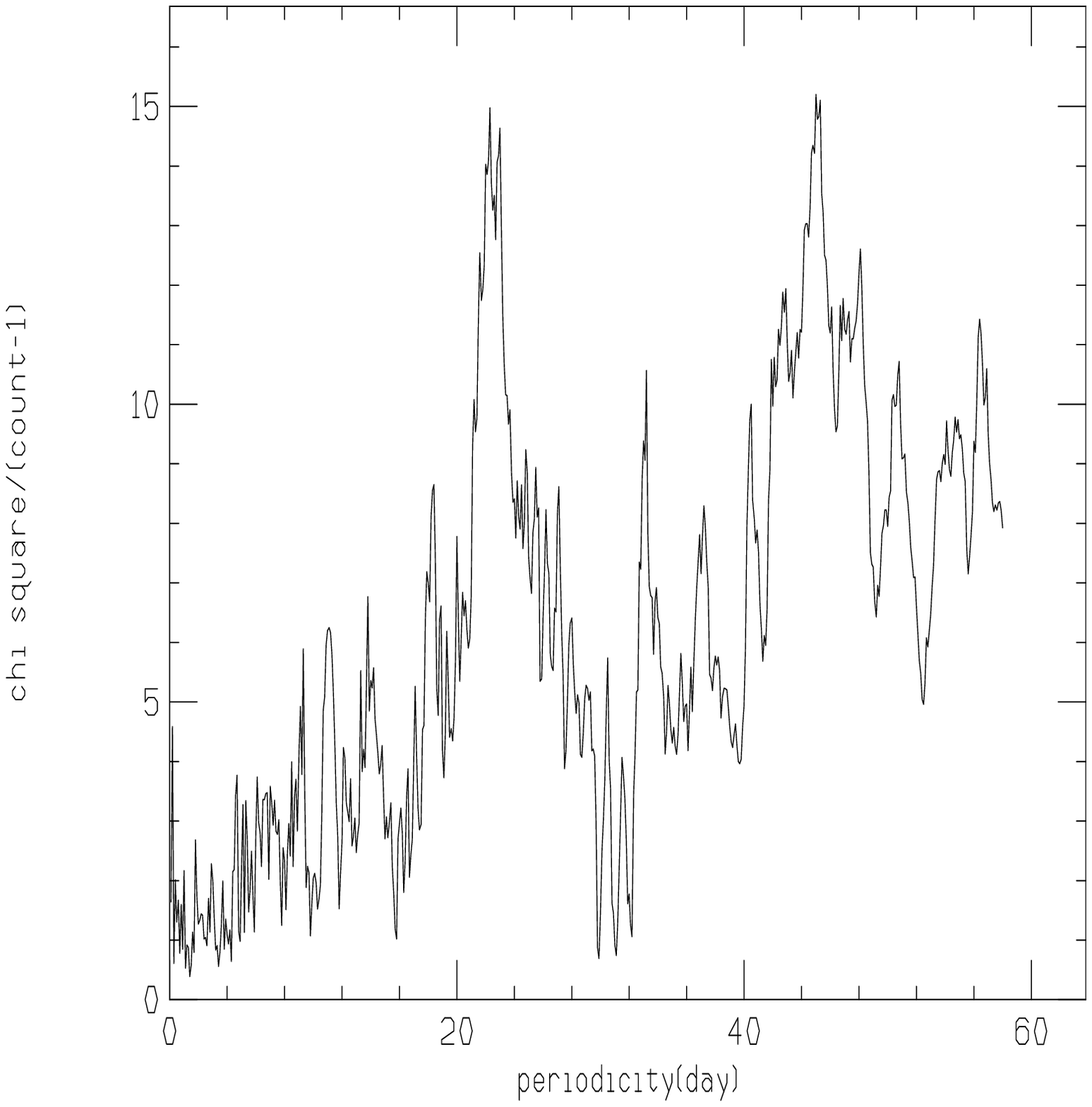}
 \end{center}
\caption{The $\chi^2$/d.o.f diagram with epoch folding method for HEGRA data(left) and ASM data(right)}
\end{figure}

\begin{figure}[p]
 \begin{center}
\includegraphics[width=6cm, height=6cm]{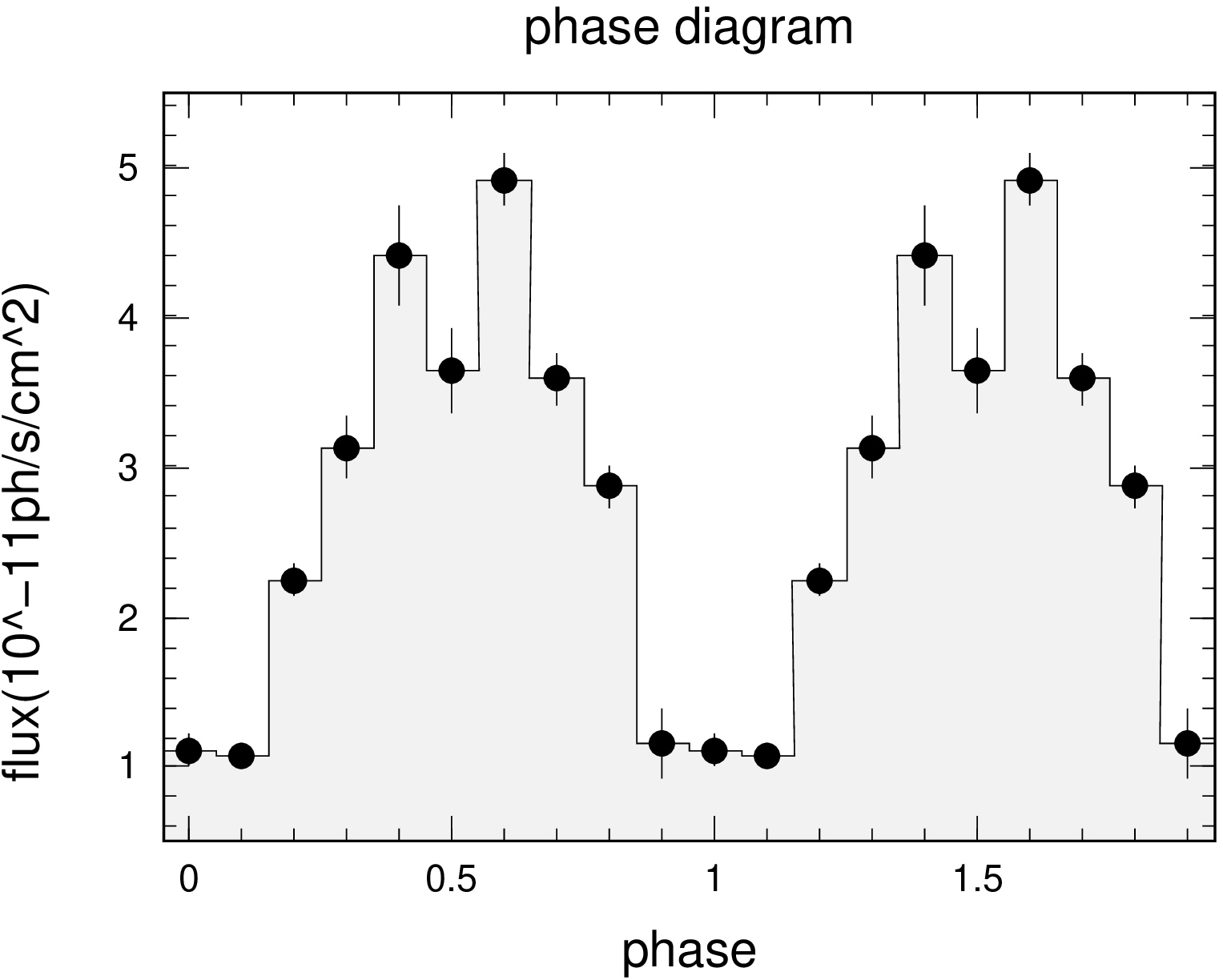}
\includegraphics[width=6cm, height=6cm]{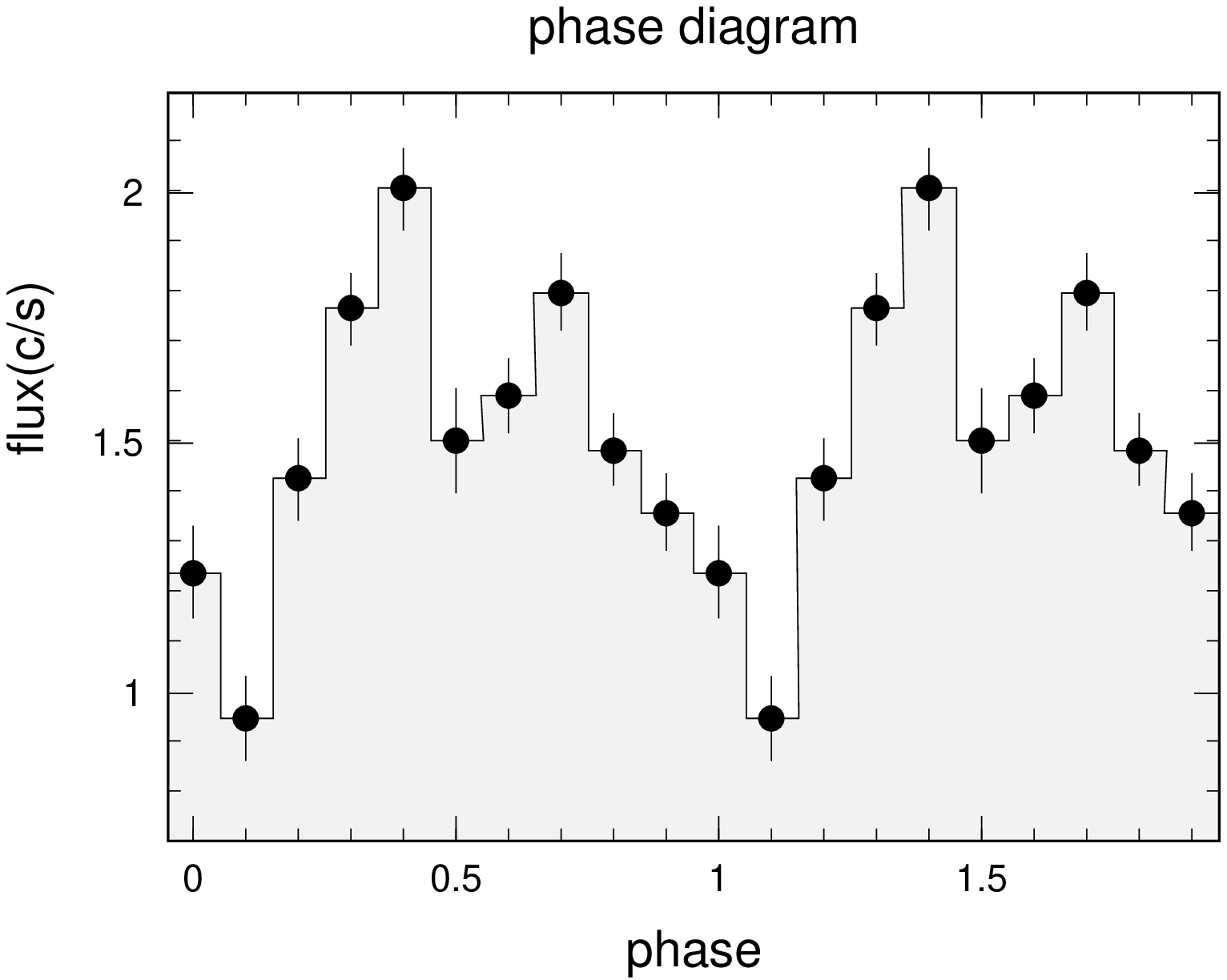}
 \end{center}
\caption{The phase diagram of HEGRA data(left) and ASM data(right) for a 22.5 day periodicity.}
\end{figure}

\begin{figure}[p]
 \begin{center}
\includegraphics[width=6cm, height=6cm]{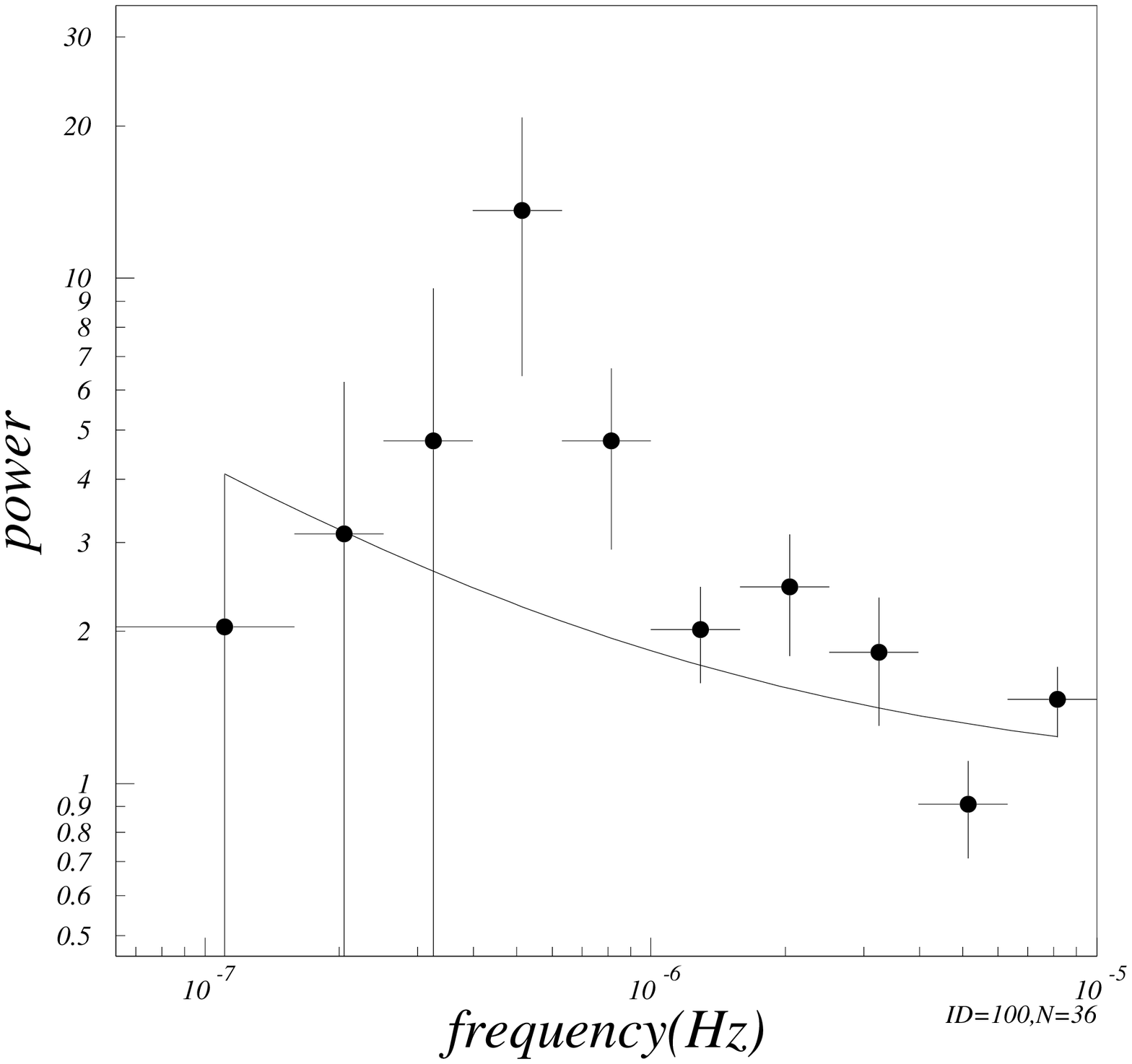}
\includegraphics[width=6cm, height=6cm]{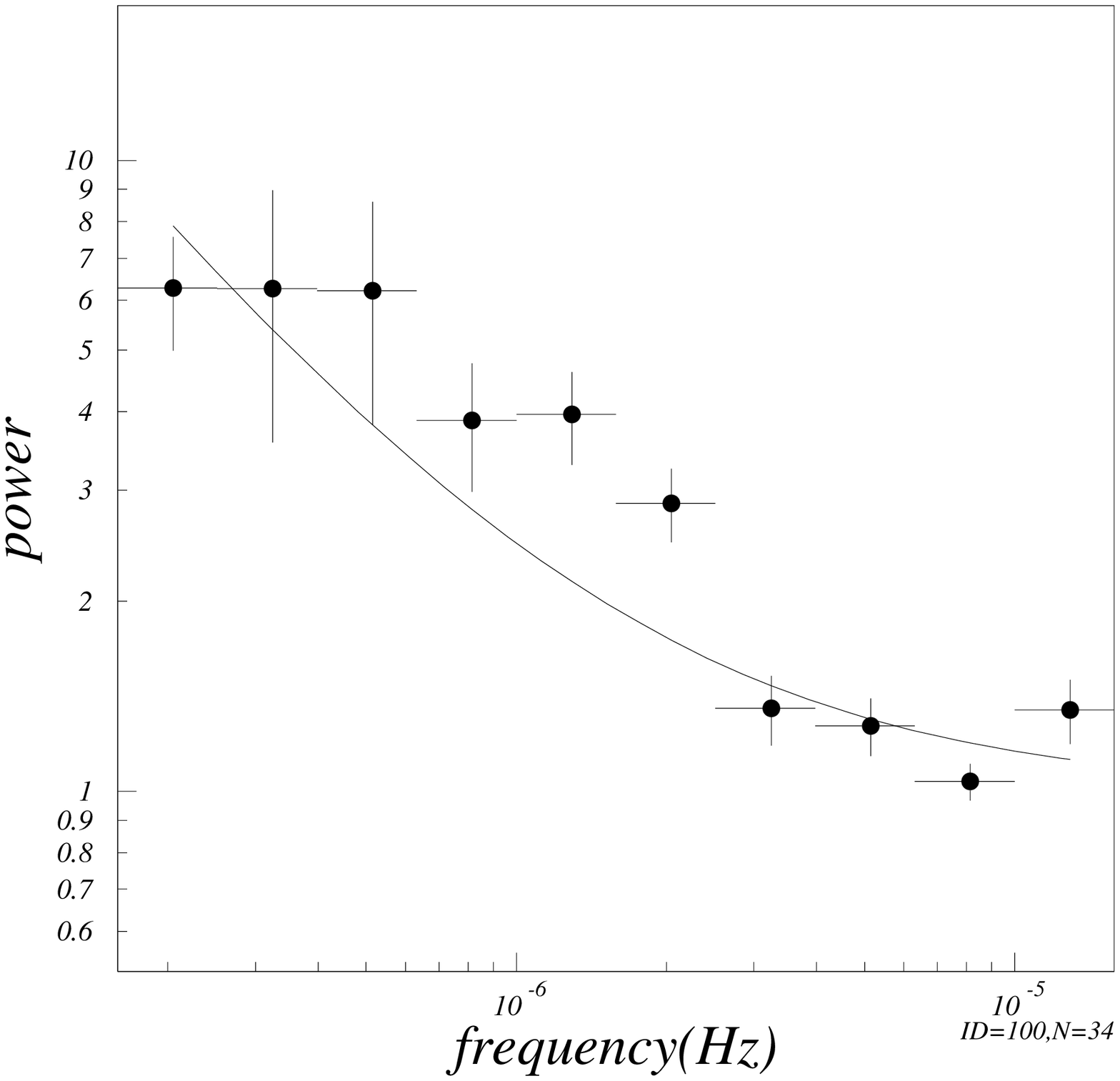}
\includegraphics[width=6cm, height=6cm]{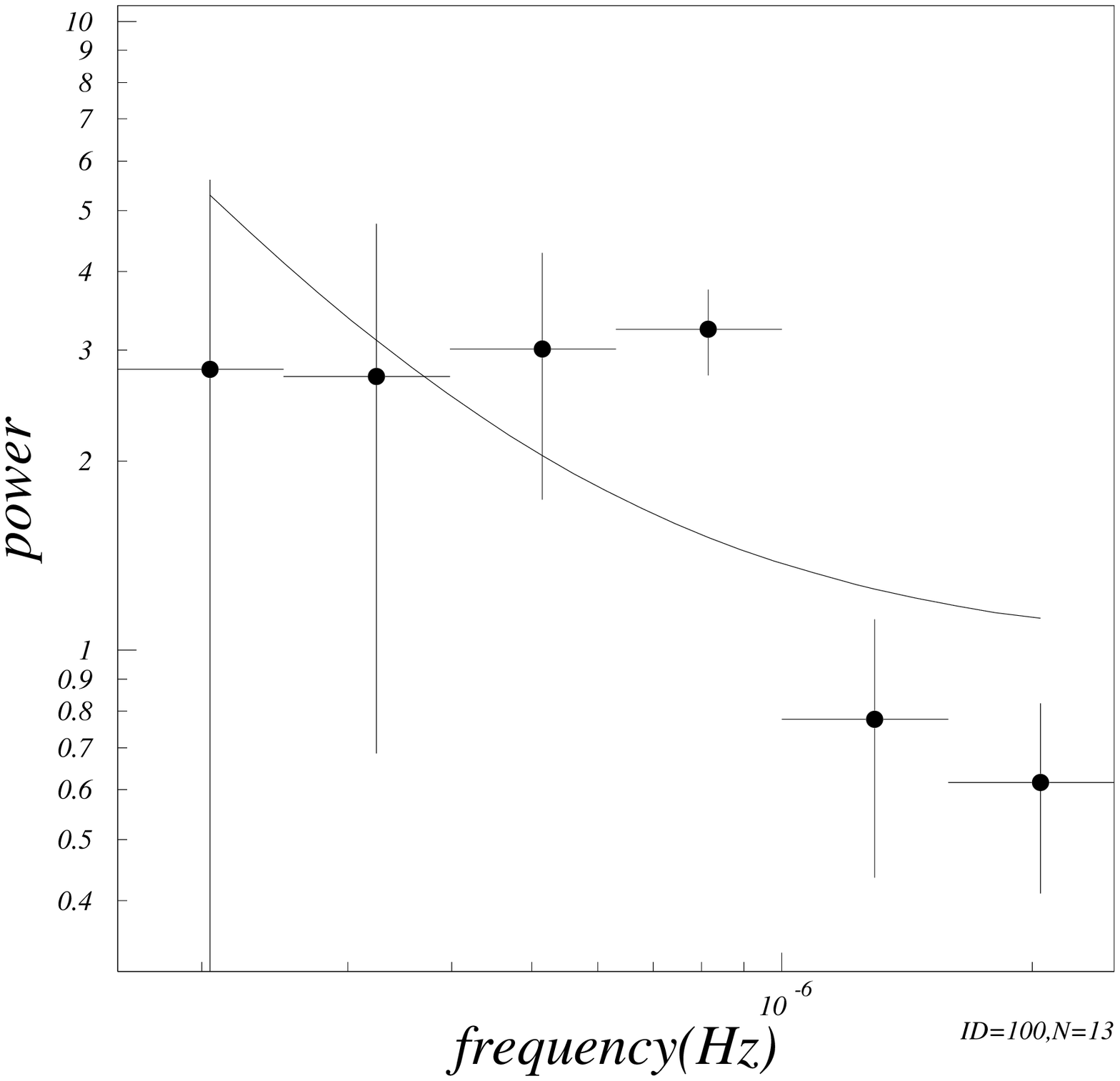}
 \end{center}
\caption{The binned power spectra for HEGRA data(Top left), ASM data(Top right) and Utah TA data(bottom). A power is fitted with a 1/f noise model in a from of $P=1+\alpha \times f^{-\beta}$(solid line). Here, $\alpha, \beta$ are constants and $f$ is a frequency. Data point $5\times10^{-7}$Hz in a HEGRA binned power spectra is removed as a periodicity when it is fitted.}
\end{figure}

\begin{figure}[p]
\begin{center}
\includegraphics[width=6cm, height=6cm]{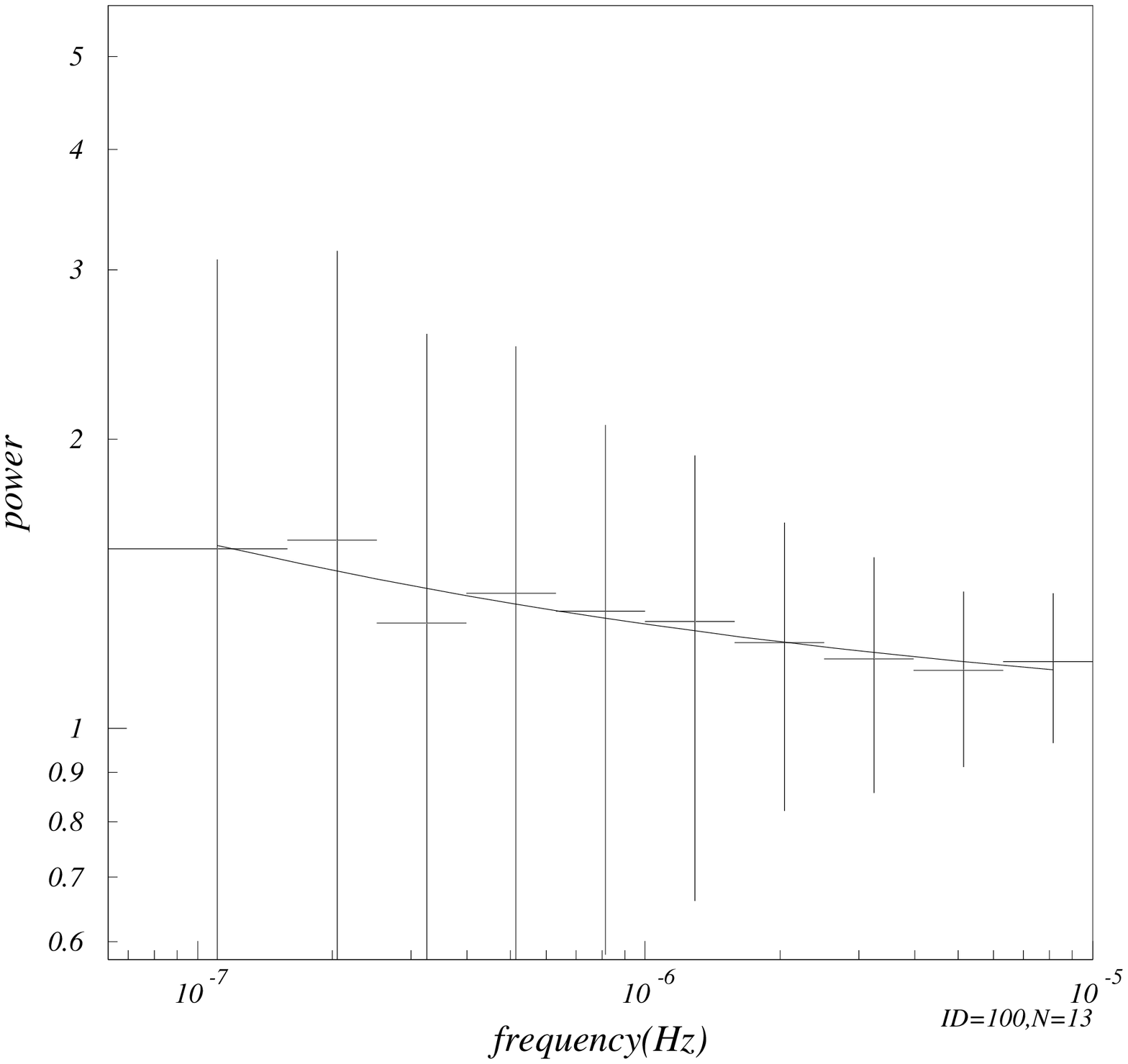}
\includegraphics[width=6cm, height=6cm]{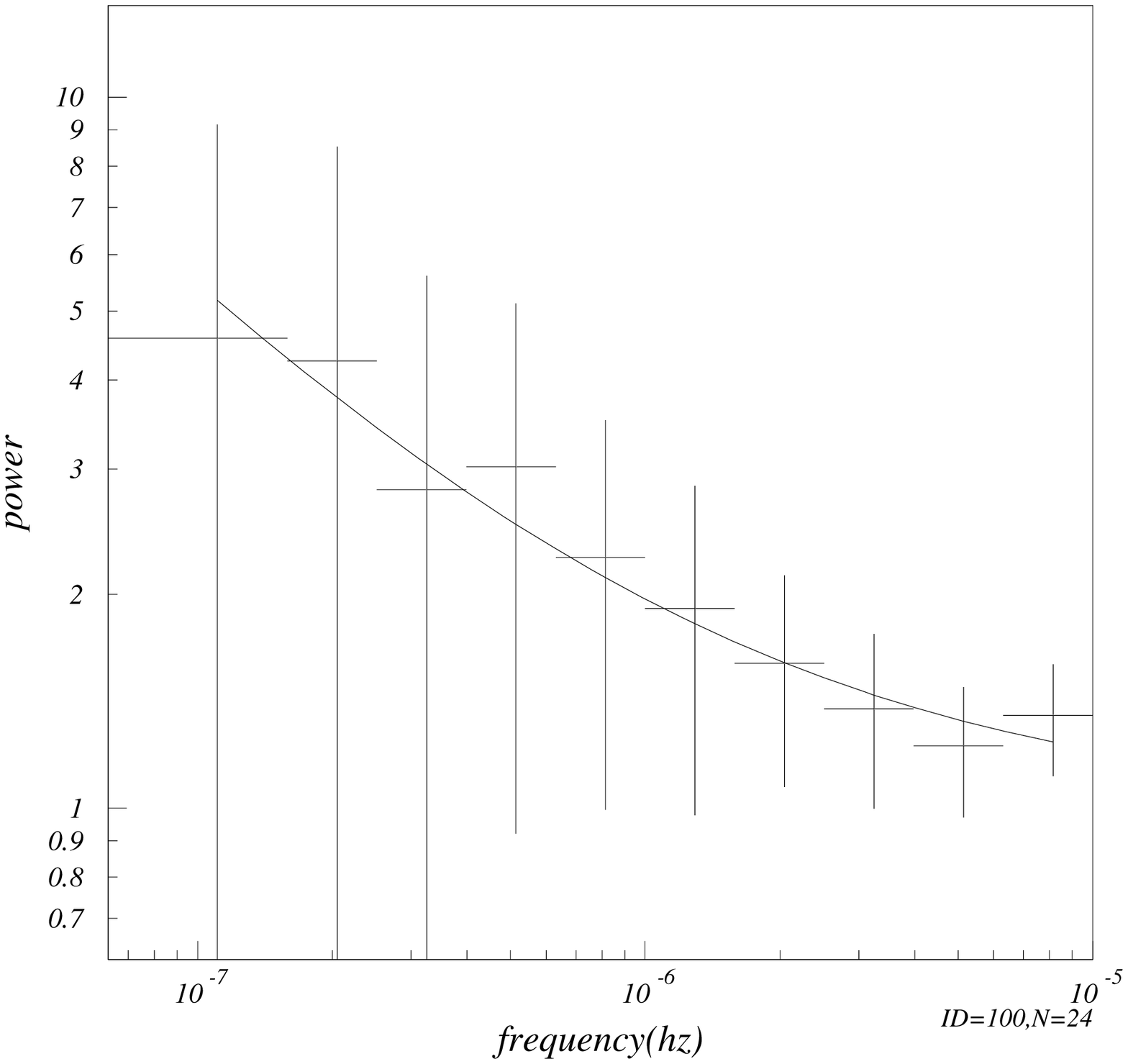}
\includegraphics[width=6cm, height=6cm]{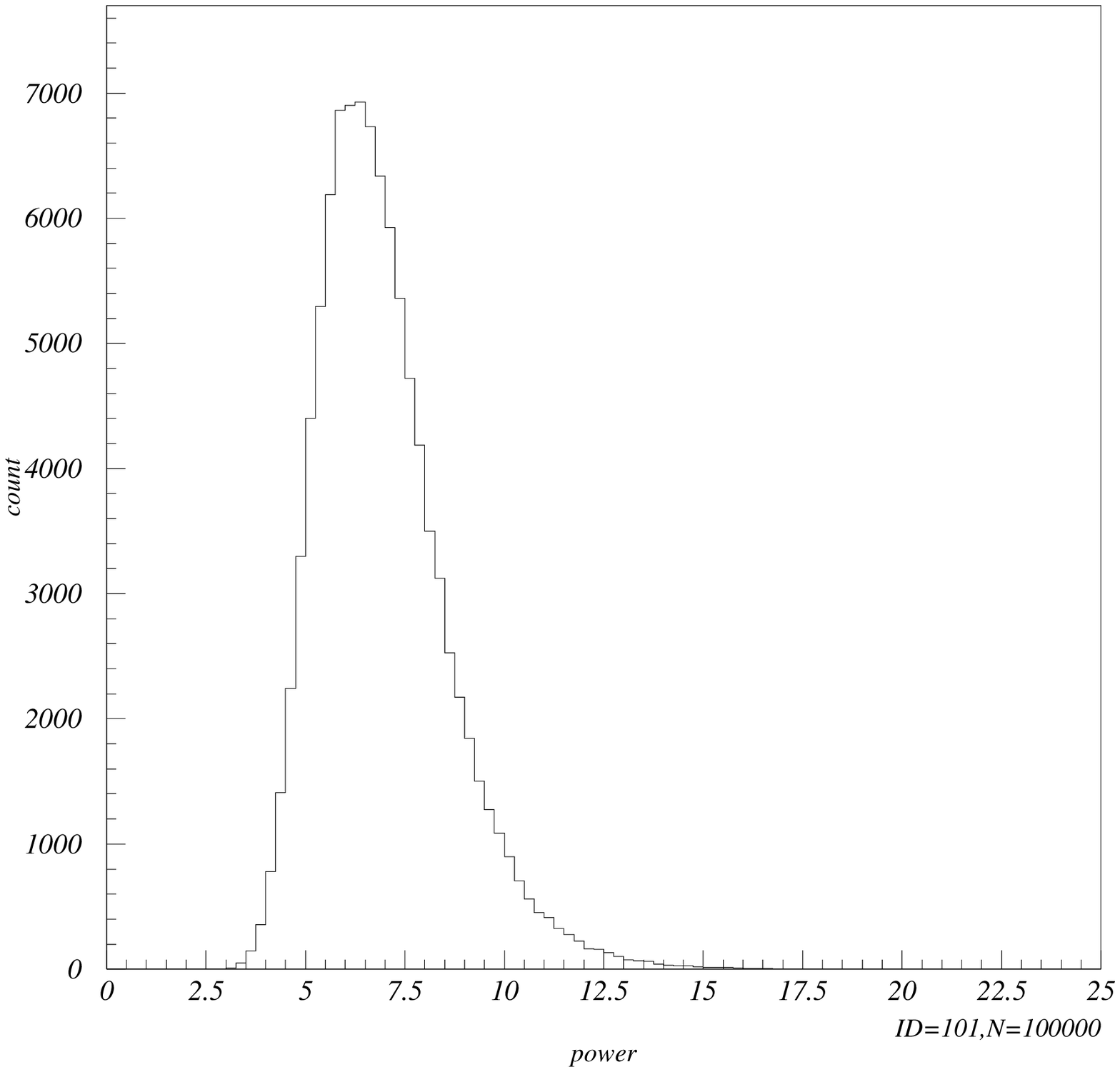}
\includegraphics[width=6cm, height=6cm]{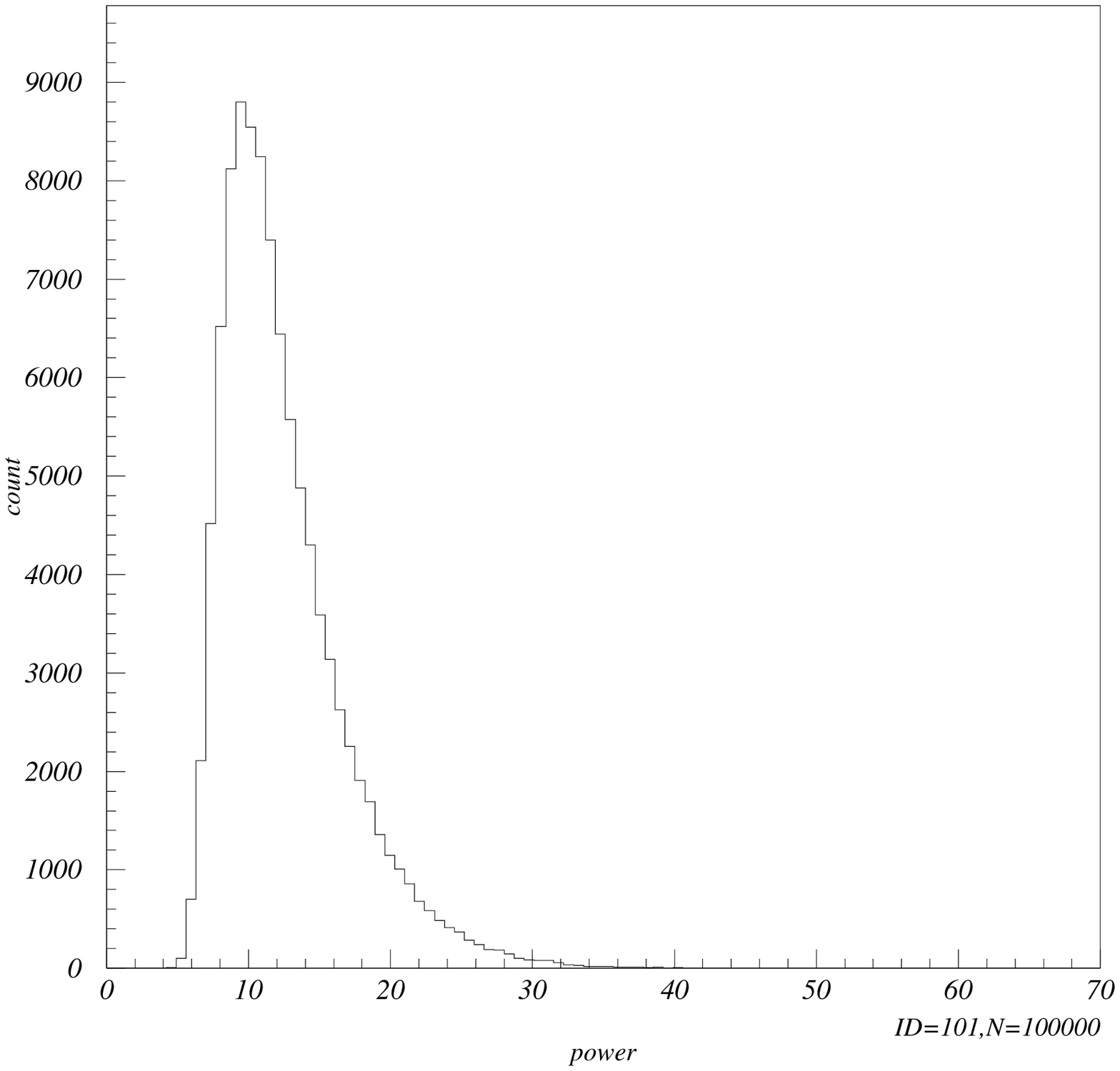}
\end{center}
\caption{The simulated binned power spectra of two extreme conditions [top left:case 1, top right:case 2] for HEGRA data. A power is fitted with a 1/f noise model in a from of $P=1+\alpha \times f^{-\beta}$(solid line). Here, $\alpha, \beta$ are constants and $f$ is a frequency. The distribution of a maximum power[bottom left:case1, bottom right:case 2] for HEGRA data}
\end{figure}

\begin{figure}[p]
\begin{center}
\includegraphics[width=6cm, height=6cm]{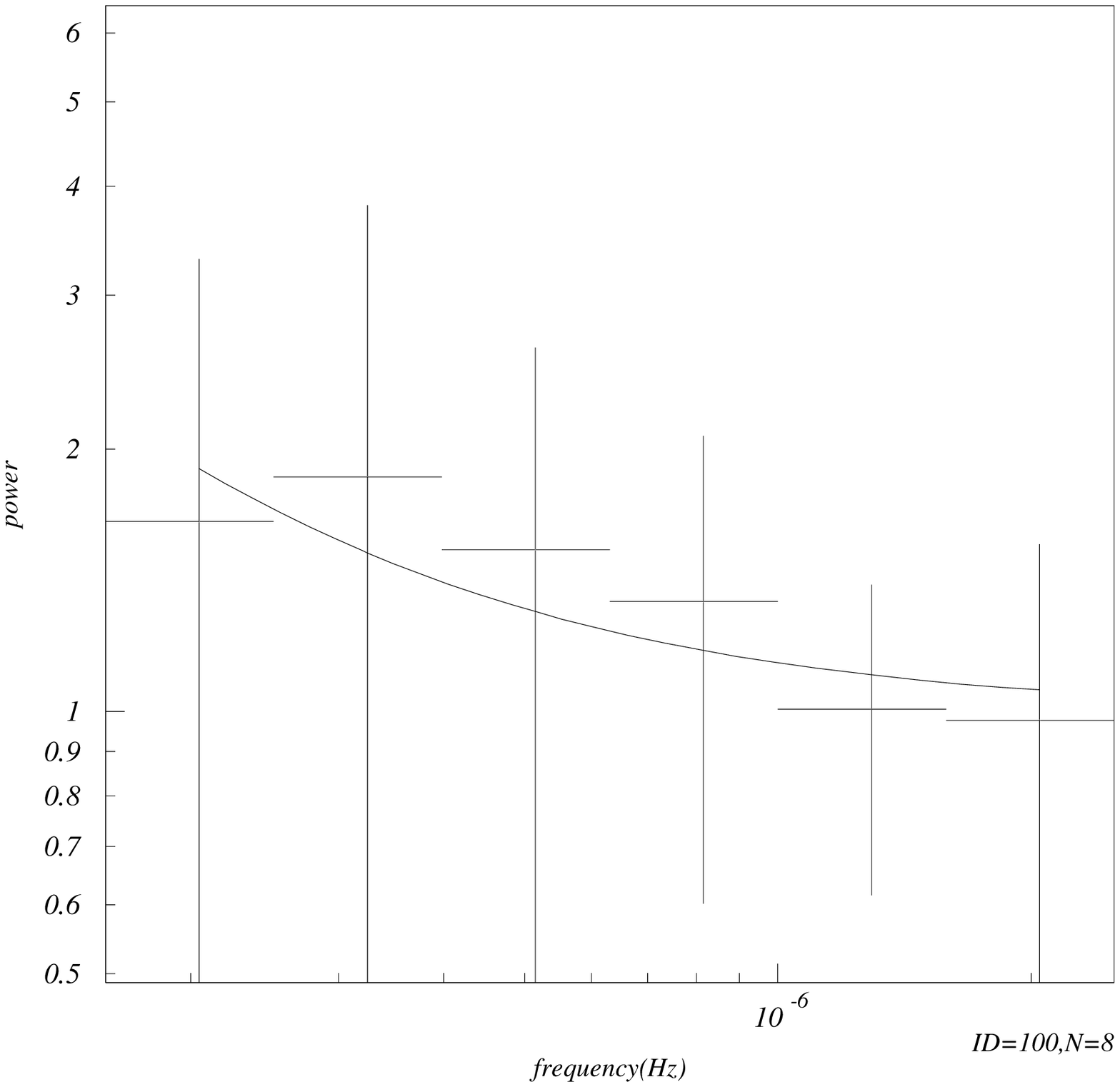}
\includegraphics[width=6cm, height=6cm]{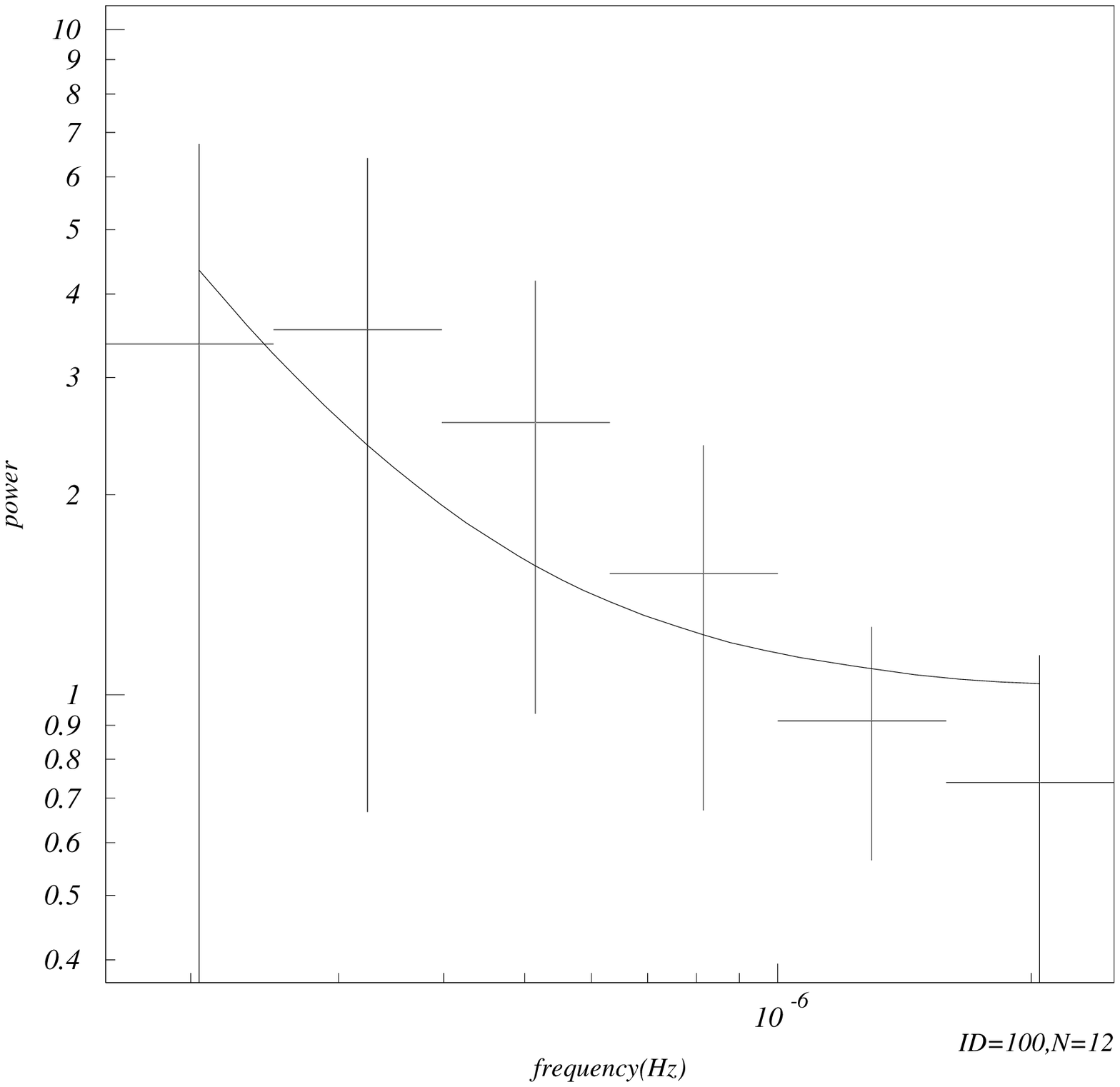}
\includegraphics[width=6cm, height=6cm]{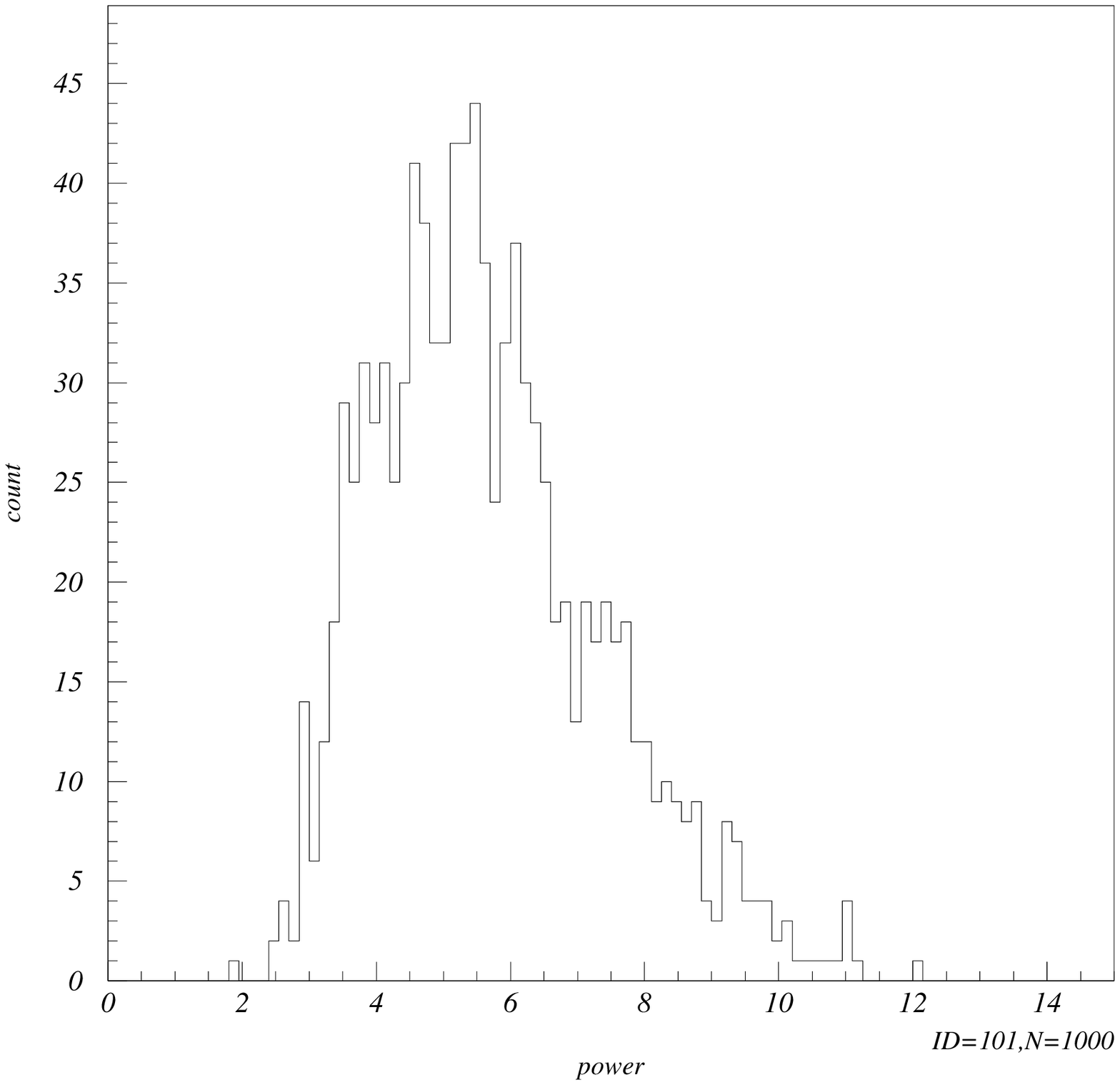}
\includegraphics[width=6cm, height=6cm]{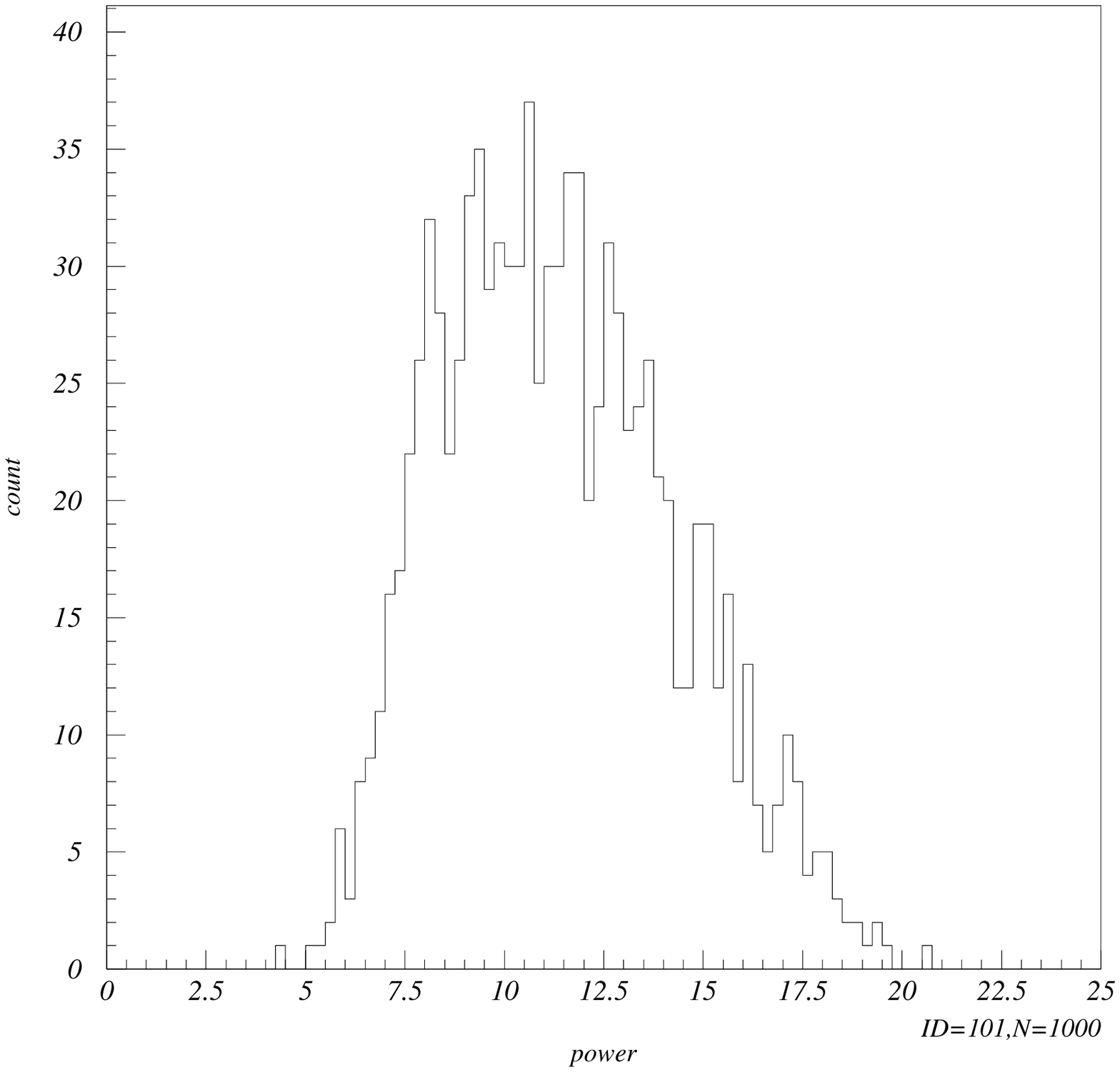}
\end{center}
\caption{The simulated binned power spectra of two extreme conditions [top left:case 1, top right:case 2] for Utah TA data and distribution of a maximum power[bottom left:case1, bottom right:case 2] for Utah TA data}
\end{figure}

\begin{figure}[p]
\begin{center}
\includegraphics[width=6cm, height=6cm]{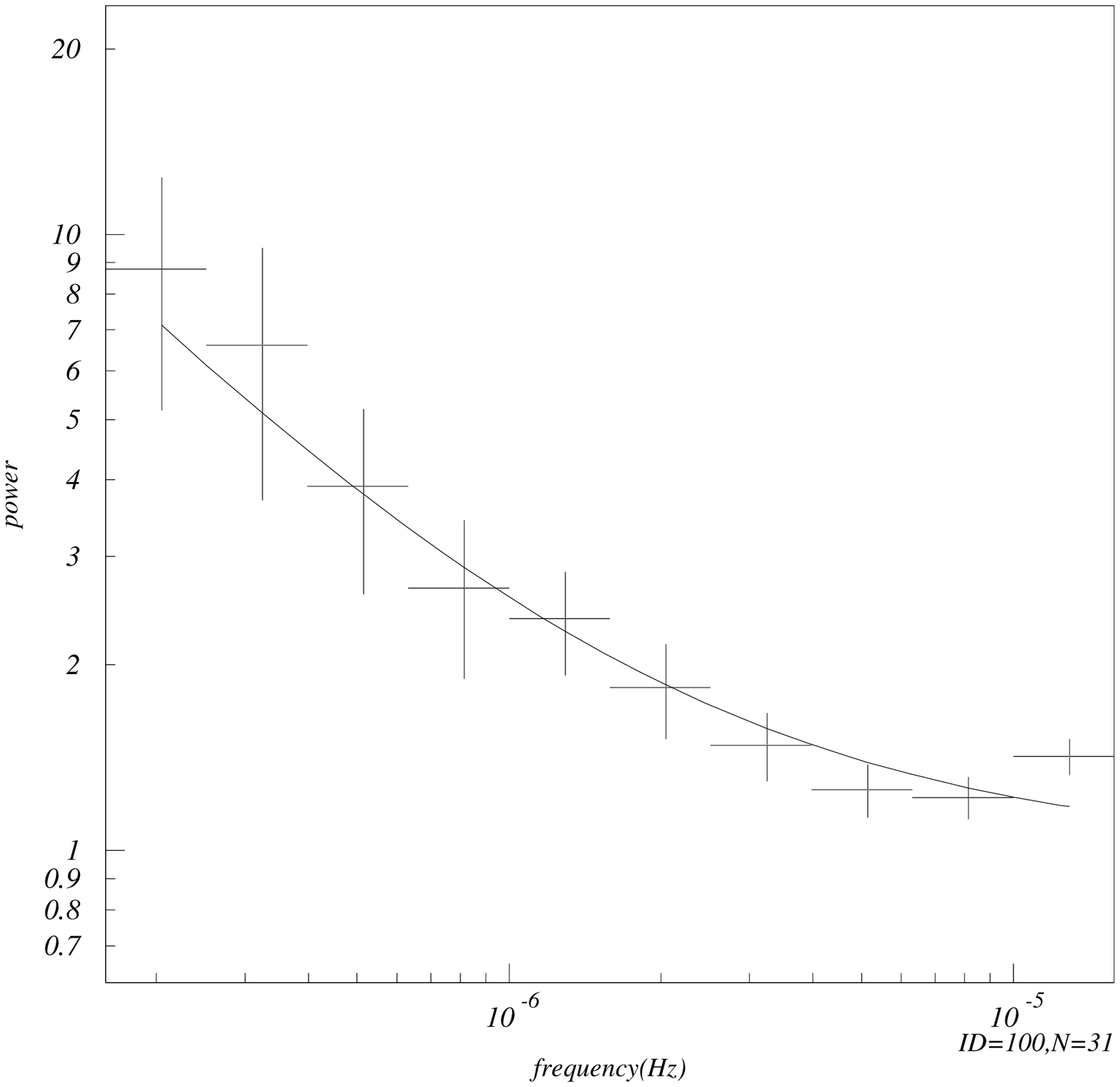}
\includegraphics[width=6cm, height=6cm]{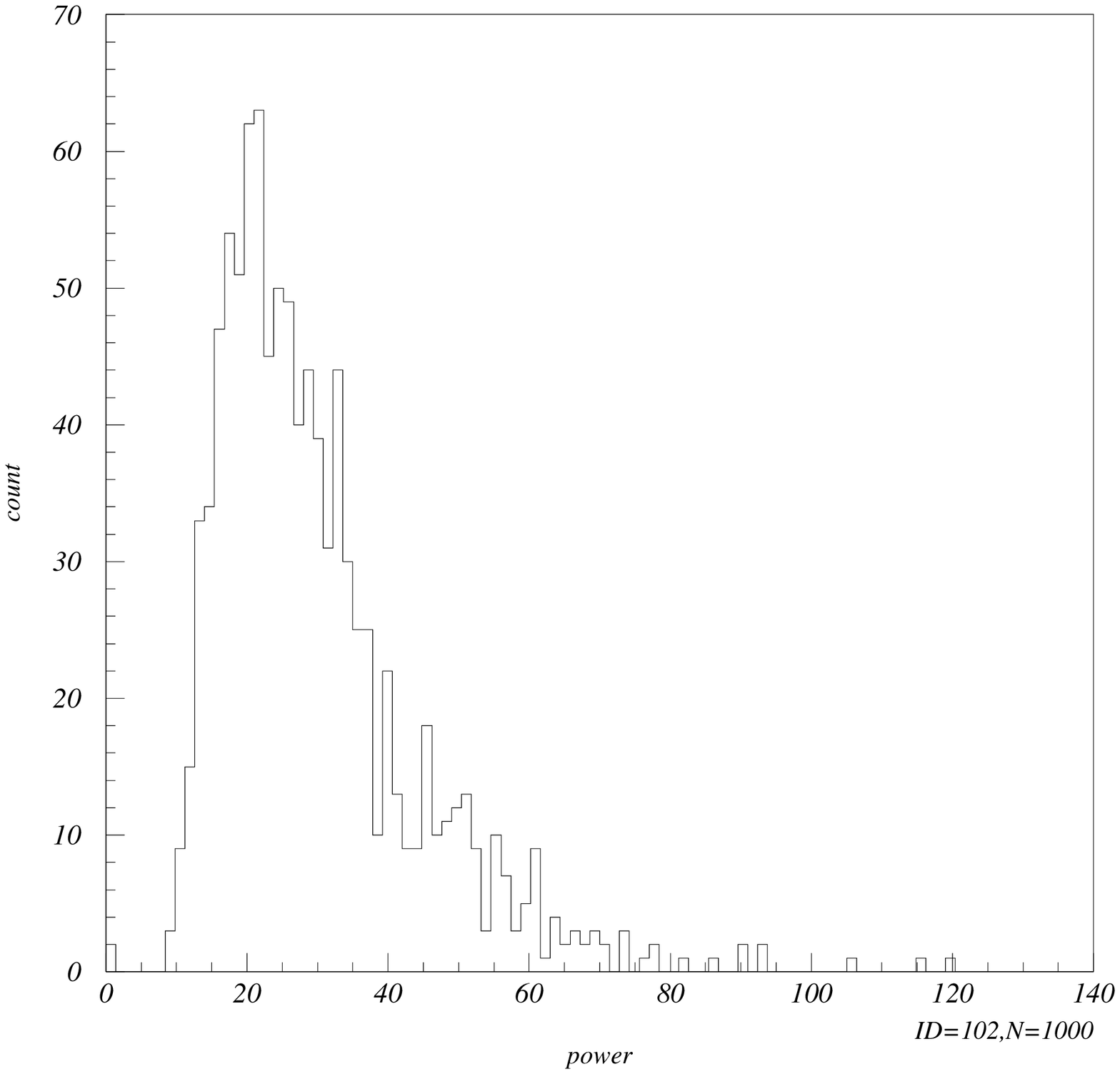}
\end{center}
\caption{The simulated binned power spectra of an extreme condition [left:case 1=case 2] and distribution of a maximum power[right:case1=case 2] for ASM data}
\end{figure}

\begin{figure}[p]
 \begin{center}
\includegraphics[width=6cm, height=6cm]{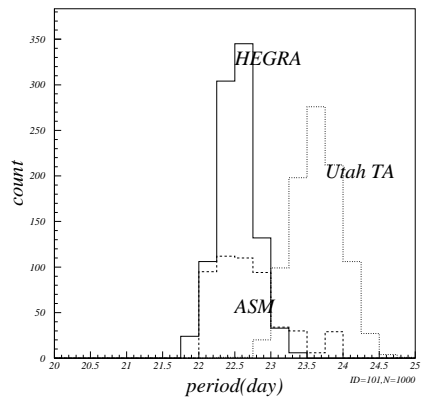}
 \end{center}
\caption{The distribution of periodicities deduced from simulation data for HEGRA data, ASM data and Utah TA data.}
\end{figure}

%\begin{figure}[p]
%\begin{center}
%\epsfxsize=6cm \epsfysize=6cm \epsfbox{./figure17.eps}
%\epsfxsize=6cm \epsfysize=6cm \epsfbox{./figure18.eps}
% \end{center}
%\caption{The power spectra for HEGRA data in MJD 50545-50603(left) and MJD 50603-50661(right).}
%\end{figure}

\begin{figure}[p]
 \begin{center}
\includegraphics[width=6cm, height=6cm]{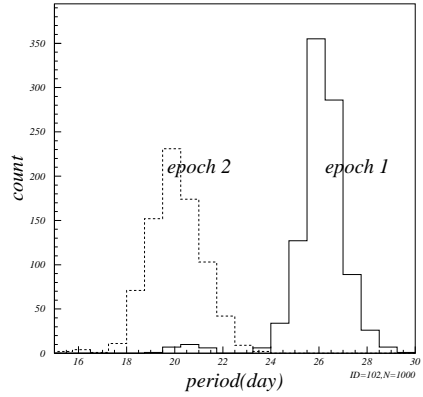}
 \end{center}
\caption{The distribution of periodicities deduced from simulation data for HEGRA data in MJD 50545-50603(epoch 1) and MJD 50603-50661(epoch 2).}
\end{figure}

%\begin{figure}[p]
% \begin{center}
%\epsfxsize=6cm \epsfysize=6cm \epsfbox{./figure19.ps}
%\epsfxsize=6cm \epsfysize=6cm \epsfbox{./figure20.ps}
% \end{center}
%\caption{The $\chi$/d.o.f method for HEGRA data in MJD 50545-50603(left) and MJD 50603-50661(right).}
%\end{figure}

\begin{figure}[p]
 \begin{center}
\includegraphics[width=6cm, height=6cm]{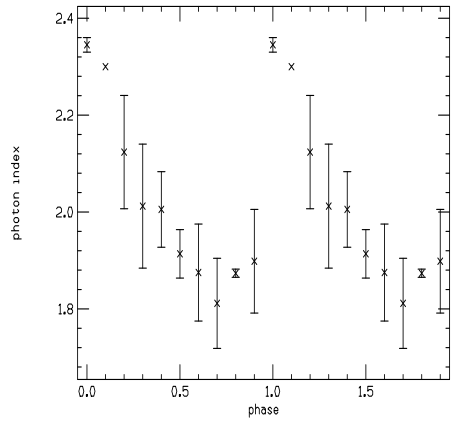}
 \end{center}
\caption{The phase diagram of a photon index with PCA data in MJD 50300-50900. The phase is folded with a 23.6 day periodicity}
\end{figure}

\rotatebox{90}{%
\begin{minipage}{17cm}
\begin{longtable}{c|ccc}
\caption{The result of a timing analysis for HEGRA, Utah TA and ASM data in 1997 for Mkn501. The error is a 1 sigma statistical error. ($A,\beta$) are fitting parameters of a 1/f noise.} \\
& HEGRA & Utah TA & ASM \\ \hline
$\sigma^2$ &6.34 ($10^{-11}$ph s$^{-1}$cm$^{-2}$)$^2$ & 0.05 (cts/min)$^2$  & 2.55 (cts/s)$^2$  \\ \hline
fourier analysis (day)  & 22.5$\pm$0.3  & 23.6$\pm$0.3  & 22.6$\pm$0.5 \\
epoch folding method(day) & 22.4   & ---   & 22.3 \\  \hline
A & (0.44$\pm0.14) \times10^{-5}$ & (0.22$\pm0.09)\times10^{-5}$ & (0.68$\pm0.09)\times10^{-5}$ \\
$\beta$ & 0.50$^{+0.16}_{-0.20}$ & 1.54$^{+0.43}_{-0.35}$ & 0.97$\pm$0.08 \\ \hline
N(Pmax$\ge$Pobs)/N$_0$(case1) & 0/10$^5$ &  296/10$^3$ & 200/10$^3$  \\ 
N(Pmax$\ge$Pobs)/N$_0$(case2) & 488/10$^5$ &  981/10$^3$ & 200/10$^3$ \\ 
chance probability & 4.88$\times10^{-3}$  & 0.981 & 0.200 \\ \hline
\end{longtable}  
\end{minipage}
}

\begin{table}
\begin{center}
\caption{A periodicity for HEGRA data with fourier analysis in two epochs. The error of periodicity is a 1 sigma statistical error.}
\begin{tabular}{c|cc}
 & MJD50545-50603 & MJD 50603-50661 \\ \hline
$\sigma^2$ ($10^{-11}$ph s$^{-1}$cm$^{-2}$)$^2$ & 5.33 & 7.29 \\
(day)  &  26.2$\pm$0.81  & 20.1$\pm$1.0 \\
%$\chi^2$/d.o.f method (day) & 20.9 & 25.6\\ 
\end{tabular}
\end{center}
\end{table}  

\end{document}